\documentclass[twocolumn]{aastex702}

\usepackage{amsmath}
\usepackage{empheq}
\usepackage{mathrsfs}
\usepackage{textcomp}
\usepackage{enumitem}   
\usepackage{gensymb}
\usepackage{hyperref}
\usepackage{graphicx}
\usepackage[caption=false]{subfig}
\usepackage{multirow}
\usepackage{longtable}
\usepackage{ulem}
\usepackage[flushleft]{threeparttable}
\usepackage{booktabs} 
\usepackage{CJK}
\hypersetup{colorlinks, linkcolor={blue}, citecolor={blue}, urlcolor={blue}} 

\newcommand{\galfit}{\texttt{GALFIT}}
\newcommand{\uprime}{\mbox{$u^\prime$}}

\definecolor{DarkOrange}{RGB}{204, 85, 0}
\definecolor{LincolnGreen}{RGB}{17, 102, 0}

\usepackage{xspace}

\newcommand{\sersic}{{S\'{e}rsic}\xspace}

\begin{document}
\begin{CJK*}{UTF8}{gbsn}
\title{AT\,2024dky as an Off-nuclear Tidal Disruption Event Candidate}

\correspondingauthor{Yuhan Yao}
\email[show]{yyao@pku.edu.cn}

\author[0000-0001-6747-8509]{Yuhan Yao}
\affiliation{Kavli Institute for Astronomy and Astrophysics, Peking University, Beijing 100871, China}
\email{yyao@pku.edu.cn}

\author[0000-0002-7706-5668]{Ryan Chornock}
\affiliation{Department of Astronomy, University of California, Berkeley, CA 94720-3411, USA}
\affiliation{Berkeley Center for Multi-messenger Research on Astrophysical Transients and Outreach (Multi-RAPTOR), University of California, Berkeley, CA 94720-3411, USA}
\email{chornock@berkeley.edu}

\author[0000-0003-0466-3779]{Itai Sfaradi}
\affiliation{Department of Astronomy, University of California, Berkeley, CA 94720-3411, USA}
\affiliation{Berkeley Center for Multi-messenger Research on Astrophysical Transients and Outreach (Multi-RAPTOR), University of California, Berkeley, CA 94720-3411, USA}
\email{itai.sfaradi@berkeley.edu}

\author[0000-0002-4569-9009]{Jinyi Shangguan}
\affiliation{Kavli Institute for Astronomy and Astrophysics, Peking University, Beijing 100871, China}
\email{shangguan@pku.edu.cn}

\author[0000-0003-4768-7586]{Raffaella Margutti}
\affiliation{Department of Astronomy, University of California, Berkeley, CA 94720-3411, USA}
\affiliation{Berkeley Center for Multi-messenger Research on Astrophysical Transients and Outreach (Multi-RAPTOR), University of California, Berkeley, CA 94720-3411, USA}
\affiliation{Department of Physics, University of California, 366 Physics North MC 7300, Berkeley, CA 94720, USA}
\email{rmargutti@berkeley.edu}

\author[0000-0003-2434-0387]{Robert Stein}
\affiliation{Deutsches Zentrum f\"{u}r Astrophysik, Postplatz 1, 02826, G\"{o}rlitz, Germany}
\email{robert.stein@dzastro.de}

\author[0000-0002-0587-6680]{Elena Maria Rossi}
\affiliation{Leiden Observatory, Leiden University, PO Box 9513, 2300 RA, Leiden, Netherlands}
\email{emr@strw.leidenuniv.nl}

\author[0009-0008-8549-8069]{Yijia Zhang}
\affiliation{Kavli Institute for Astronomy and Astrophysics, Peking University, Beijing 100871, China}
\email{zhangyij26@pku.edu.cn}

\author[0000-0001-8701-2116]{Zijian Zhang (张子健)}
\affiliation{Department of Physics, The University of Hong Kong, Pokfulam Road, Hong Kong, China}
\affiliation{The Hong Kong Institute for Astronomy and Astrophysics, The University of Hong Kong, Pokfulam Road, Hong Kong, China}
\email{astrozzj@connect.hku.hk}

\author[0000-0002-9589-5235]{Jane Lixin Dai}
\affiliation{Department of Physics, The University of Hong Kong, Pokfulam Road, Hong Kong, China}
\affiliation{The Hong Kong Institute for Astronomy and Astrophysics, The University of Hong Kong, Pokfulam Road, Hong Kong, China}
\email{lixindai@hku.hk}

\author[0000-0002-1568-7461]{Wenbin Lu}
\affiliation{Department of Astronomy, University of California, Berkeley, CA 94720-3411, USA}
\affiliation{Theoretical Astrophysics Center, University of California, Berkeley, CA 94720, USA}
\affiliation{Berkeley Center for Multi-messenger Research on Astrophysical Transients and Outreach (Multi-RAPTOR), University of California, Berkeley, CA 94720-3411, USA}
\email{wenbinlu@berkeley.edu}


\author[0000-0001-8018-5348]{Eric C. Bellm}
\affiliation{DIRAC Institute, Department of Astronomy, University of Washington, 3910 15th Avenue NE, Seattle, WA 98195, USA}
\email{ecbellm@uw.edu}

\author[0000-0002-7777-216X]{Joshua S. Bloom}
\affiliation{Department of Astronomy, University of California, Berkeley, CA 94720-3411, USA}
\affiliation{Berkeley Center for Multi-messenger Research on Astrophysical Transients and Outreach (Multi-RAPTOR), University of California, Berkeley, CA 94720-3411, USA}
\affiliation{Lawrence Berkeley National Laboratory, 1 Cyclotron Road, MS 50B-4206, Berkeley, CA 94720, USA}
\email{jbloom@astro.berkeley.edu}

\author[0000-0003-0528-202X]{Collin T. Christy}
\affiliation{Steward Observatory, University of Arizona, 933 North Cherry Avenue, Tucson, AZ85721-0065, USA}
\email{collinchristy@arizona.edu}

\author[0000-0002-5884-7867]{Richard Dekany}
\affiliation{Caltech Optical Observatories, California Institute of Technology, Pasadena, CA 91125, USA}
\email{rgd@astro.caltech.edu}

\author[0000-0002-3168-0139]{Matthew J. Graham}
\affiliation{Cahill Center for Astronomy and Astrophysics, 
    California Institute of Technology, MC 249-17, 
    1200 E California Boulevard, Pasadena, CA 91125, USA}
\email{mjg@caltech.edu}

\author[0000-0002-5619-4938]{Mansi M. Kasliwal}
\affiliation{Cahill Center for Astronomy and Astrophysics, 
	California Institute of Technology, MC 249-17, 
	1200 E California Boulevard, Pasadena, CA 91125, USA}
\email{mansi@astro.caltech.edu}

\author[0000-0003-2451-5482]{Russ R. Laher}
\affiliation{IPAC, California Institute of Technology, 1200 E. California Blvd, Pasadena, CA 91125, USA}
\email{laher@ipac.caltech.edu}

\author[0000-0003-2242-0244]{Ashish~A.~Mahabal}
\affiliation{Division of Physics, Mathematics and Astronomy, California Institute of Technology, Pasadena, CA 91125, USA}
\affiliation{Center for Data Driven Discovery, California Institute of Technology, Pasadena, CA 91125, USA}
\email{aam@astro.caltech.edu}

\author[0000-0003-1227-3738]{Josiah Purdum}
\affiliation{Caltech Optical Observatories, California Institute of Technology, Pasadena, CA 91125, USA}
\email{jpurdum@caltech.edu}

\author[0000-0001-7648-4142]{Ben Rusholme}
\affiliation{IPAC, California Institute of Technology, 1200 E. California Blvd, Pasadena, CA 91125, USA}
\email{rusholme@ipac.caltech.edu}

\begin{abstract} 
Through a retrospective search for tidal disruption events (TDEs) in the ZTF archives agnostic to their location within their host galaxies, we identify AT\,2024dky/ZTF24aagksvd as a promising off-nuclear TDE candidate. 
At $z=0.0592$, it lies at a projected offset of $3.72\arcsec$ (4.26\,kpc) from the center of a massive ($M_{\rm gal}\sim10^{11}\,M_\odot$), quiescent galaxy. 
We present early-time ZTF and ATLAS optical photometry and late-time follow-up observations with Keck-I/LRIS, VLA, EP, and Chandra. 
With $M_{g,\rm peak}=-18.50$\,mag, AT\,2024dky exhibits persistently blue color ($g-r\sim-0.1$\,mag) through 80\,d after peak. 
At $\delta t=858$\,d, Keck reveals $M_{u'}=-13.07\pm0.09$~mag and $M_R > -14.15$\,mag. 
At $\delta t=866$\,d, VLA detects the transient at $\nu L_{\nu,10\,\rm GHz}=(1.26\pm0.17)\times10^{38}\,\mathrm{erg\,s^{-1}}$.
The combination of early photometric evolution, detected late-time radio and UV emission strongly favor a TDE interpretation.
The properties of its host galaxy also bear remarkable similarities to the parent galaxies of known off-nuclear TDEs. 
Based on the low values of peak-time and late-time UV/optical luminosities, we favor a black hole mass of $\sim 10^5$--$10^6\,M_\odot$.
Together, these observations strongly favor an off-nuclear TDE.
\end{abstract}


\section{Introduction}

Massive black holes (MBHs) residing outside galactic nuclei, including both supermassive black holes (SMBHs; $M_{\rm BH}\gtrsim 10^{5}\,M_\odot$) and intermediate-mass black holes (IMBHs; $10^2\lesssim M_{\rm BH}\lesssim 10^{5}\,M_\odot$), can arise through several formation channels. During galaxy minor mergers, tidal stripping can leave the central MBH of the secondary galaxy embedded in a stripped nucleus, while long dynamical friction timescales delay its migration toward the center of the parent galaxy \citep{Dosopoulou2017,Kelley2017}. MBHs can also be ejected from galactic nuclei by gravitational-wave recoil following black hole coalescence or by gravitational slingshot interactions involving multiple black holes \citep{Blecha2016,Hoffman2007,Ryu2018}. Off-nuclear IMBHs may also form in dense stellar clusters through runaway stellar collisions \cite{PortegiesZwart2004,GonzalezPrieto2024}. The abundance, spatial distribution, and stellar environments of these off-nuclear MBHs provide constraints on black hole formation, the accretion and disruption of satellite galaxies, and the efficiency with which MBHs pair and merge. They therefore offer a complementary perspective on the connection between black hole growth and galaxy assembly beyond that provided by central MBHs.

Direct evidence of the existence of off-nuclear MBHs come from dynamical detections in a few star clusters \citep{Seth2014, Ahn2018, Afanasiev2018, Pechetti2022, Taylor2025}. 
However, this technique requires resolving the sphere of influence and is therefore largely restricted to very nearby ($\lesssim 20$\,Mpc) systems.

At larger distances, off-nuclear MBHs have been searched for as offset active galactic nuclei (AGNs) or AGN pairs (either dual or binary). 
A literature compilation covering studies published through 2020 contains 156 confirmed dual AGNs and one confirmed binary AGN \citep{Pfeifle2025}.
Complementary searches have identified individual spatially offset AGNs and larger samples of candidates in merging galaxies \citep[e.g.,][]{Barrows2016,Barrows2025}.
However, AGN searches preferentially reveal MBHs with sufficient gas to sustain persistent accretion.

Tidal disruption events (TDEs) provide a complementary approach: the disruption of a star can temporarily illuminate an otherwise dormant black hole in a gas-poor, quiescent environment, making it detectable over a much larger volume. This method potentially favors MBHs embedded in dense stellar systems that can supply stars and replenish the loss cone \citep{Chang2025}. Nevertheless, theoretical studies suggest that off-nuclear black holes could contribute a significant fraction of the TDE population. Recoiling black holes alone may account for $\sim 1$\% of all disruptions \citep{Stone2012_offset}, while merger-driven wanderers could contribute substantially more \citep{Ricarte2021_tde}.

A few off-nuclear TDEs have been established. 
The first two, 3XMM\,J2150-05 and EP240222a, were identified in X-rays and are consistent with disruptions by IMBHs of $10^4$--$10^5\,M_\odot$. Whereas 3XMM\,J2150-05 was discovered through archival observations \citep{Lin2018_3XMMJ2150}, EP240222a was detected in real time, enabling prompt multi-wavelength follow-up \citep{Jin2025}. The remaining three events were discovered optically by the Zwicky Transient Facility (ZTF; \citealt{Graham2019,Bellm2019b,Dekany2020}). 
TDE\,2024tvd initially passed a nuclear-transient selection, but its significant ($\sim 1\arcsec$) offset from the host nucleus, together with its low peak luminosity relative to nuclear TDEs in comparably massive galaxies, motivated high-resolution follow-up, which confirmed its off-nuclear nature \citep{Yao2025_24tvd}. 
This discovery motivated complementary efforts to identify off-nuclear TDEs: real-time searches that relax the nuclear-location requirement lead to TDE\,2025abcr \citep{Stein2026},
and reexamination of the known ZTF nuclear TDE sample for previously unrecognized offsets identified TDE\,2023mfm \citep{Stein2026, Li2026, Sfaradi2026}.

Other off-nuclear TDE candidates include WINGS\,J1348+26 \citep{Donato2014, Maksym2013, Maksym2014}, 1eRASS\,J142140.3-295325 \citep{Grotova2025}, and the IMBH candidates ESO\,243-49 HLX-1 \citep{Farrell2009, Lasota2011} and NGC\,6099 HLX-1 \citep{Chang2025_NGC6099}.

\begin{figure*}[htbp!]
    \centering
    \includegraphics[width=0.66\textwidth]{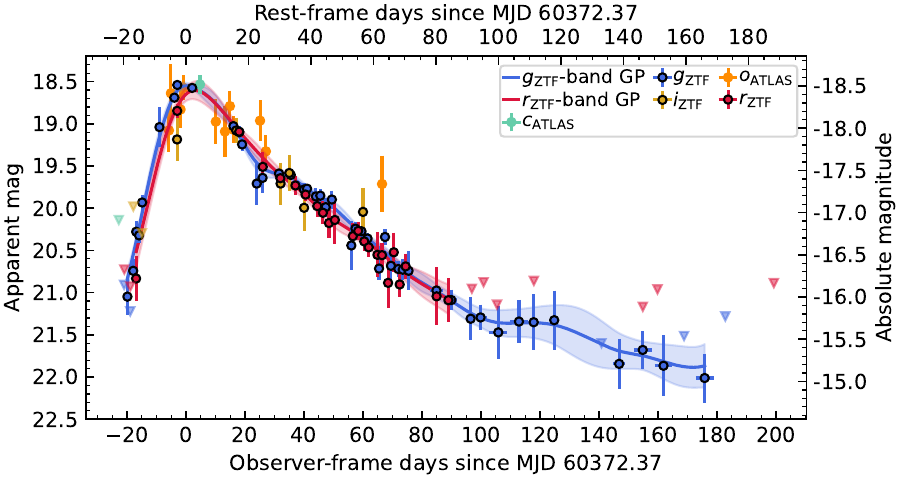}
    \includegraphics[width=0.33\textwidth]{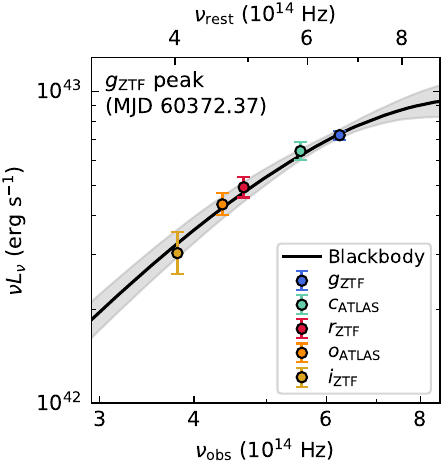}
    \caption{\textit{Left}: Optical light curve of AT\,2024dky. Solid points represent detections above 3$\sigma$; semitransparent downward triangles indicate 3$\sigma$ upper limits. The blue and red lines show the best-fit GP model and its uncertainty in $g_{\rm ZTF}$ and $r_{\rm ZTF}$ band, respectively. \textit{Right}: Optical SED of AT2024dky at the $g_{\rm ZTF}$-band peak. Colored circles show the GP-predicted photometry. The solid black curve and shaded region show the median blackbody spectrum and its 68\% credible interval.
    \label{fig:opt_lc}}
\end{figure*}

In this Letter, we report AT\,2024dky. Its displacement from the host-galaxy nucleus is unambiguous. Although the absence of early-time spectroscopy initially limits the certainty of its TDE classification, our late-time optical and radio observations, together with the resemblance of its host environment to those of known off-nuclear TDEs, strongly favor a TDE interpretation. 
We present observations of the transient and its host galaxy to assess its nature and motivate further follow-up.

We adopt a standard $\Lambda$CDM cosmology with $\Omega_{\rm M} = 0.3$, $\Omega_{\Lambda}=0.7$, and $H_0=70\,{\rm km\,s^{-1}\,Mpc^{-1}}$.
Times are given in UTC. 
Coordinates are reported in the ICRS frame and J2000 equinox. 
Magnitudes are given in the AB system.
We use the extinction law from \citet{Cardelli1989} with $R_V=3.1$, and adopt the Galactic extinction map of \citet{Schlafly2011}. 
Unless otherwise noted, uncertainties are reported at 1$\sigma$ Gaussian equivalent, and upper limits are reported at 3$\sigma$ confidence level.

\section{Discovery of AT\,2024dky} \label{sec:discovery}

To quantify off-nuclear TDE rates and identify additional candidates for demographic studies, we performed a retrospective search of ZTF transients with TDE-like photometric properties, irrespective of their location within their host galaxies. The search methodology and full candidate sample will be presented in R.~Stein, Y.~Yao, et al.\ (in preparation). We applied two independent selection methods: \texttt{tdepipe}, an empirically calibrated, physics-motivated pipeline developed by Y.~Yao that applies explicit selection cuts; and a modified version of the machine-learning classifier \texttt{tdescore}, developed by R.~Stein \citep{Stein2024}.

AT\,2024dky was first reported to the Transient Name Server (TNS) by the ZTF collaboration under the survey designation ZTF24aagksvd, based on a ZTF public-survey detection at $r_{\rm ZTF}=19.01\pm0.12$\,mag on 2024 February 29 \citep{Fremling2024}. Our retrospective search identified it as the off-nuclear TDE candidate with the brightest apparent magnitude at peak selected by \texttt{tdepipe}, and one of the highest-ranked candidates identified by \texttt{tdescore}.

\begin{figure}[htbp!]
    \centering
    \includegraphics[width=0.9\columnwidth]{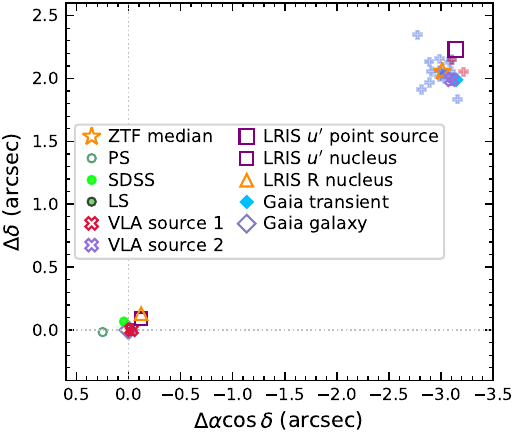}
    \caption{Locations of the host nucleus and the transient measured by different instruments. The origin of $(0, 0)$ marks the Gaia location of the host nucleus. \label{fig:offset}}
\end{figure}

AT\,2024dky generated 18 ZTF alerts, with a median position of 
${\rm R.A.}=14^{\rm h}51^{\rm m}17.599^{\rm s}$, 
${\rm decl.}=+16^{\circ}26^{\prime}01.745^{\prime\prime}$.
The transient was also detected by Gaia as Gaia24aux\footnote{\url{http://gsaweb.ast.cam.ac.uk/alerts/alert/Gaia24aux/}}, at 
${\rm R.A.}=14^{\rm h}51^{\rm m}17.590^{\rm s}$, 
${\rm decl.}=+16^{\circ}26^{\prime}01.680^{\prime\prime}$, consistent with the ZTF median position to within $0.151\arcsec$ (see Figure~\ref{fig:offset}).
We note that Gaia and ZTF alerts positions are typically accurate to $\sim0.1$\arcsec\ \citep{Hodgkin2021, Masci2019, vanvelzen2019}.

\section{Observations and Analysis} \label{sec:obs}

\subsection{Optical Light Curve} \label{subsec:optical_peak}

We obtained ZTF \citep{Masci2019, Masci2023} and ATLAS \citep{Tonry2018, Smith2020, Shingles2021} forced photometry at the Gaia transient position. 
Baseline correction was performed using the methods outlined in \citet{Yao2019}. 
We binned the ZTF and ATLAS photometry separately in each filter using the same time-dependent bin widths: 1 day before MJD 60400, 2 days between MJD 60400 and 60450, 5 days between MJD 60450 and 60475, and 7 days thereafter. Within each bin, we computed the inverse-variance-weighted mean flux and its uncertainty. When the reduced $\chi^2$ of the measurements about the weighted
mean exceeded unity, we multiplied this uncertainty by $\sqrt{\chi^2_{\rm red}}$ to account for excess scatter. The resulting light curve is shown in the left panel of Figure~\ref{fig:opt_lc}. 
Here after we use $\delta t$ to denote time since the first forced photometry $g$-band detection on 2024 February 12 (MJD 60352.5).

We model the multi-band light curve using a Gaussian process (GP; see \citealt{Aigrain2023} for a review), following the implementation described by \citet{Zhang2026_eRO}. The GP model predicts a $g$-band peak at MJD 60372.37 with $M_{g,\rm peak}=-18.50$. Using the best-fit $g$-band model, we compute that the rest-frame times from half-maximum flux to peak and from peak to half-maximum flux are $t_{1/2,\rm rise}=9.5$\,d and $t_{1/2,\rm decline}=19.6$\,d, respectively, yielding a total duration above half-maximum of $t_{1/2}=29.1$\,d.

We constructed the optical SED at $\delta t = 0$\,d using predictions from the joint GP model in the ZTF $gri$ and ATLAS $co$ bands.
We fitted a redshifted blackbody, integrating the model through each filter response. 
We sampled $\log_{10}(T_{\rm bb}/{\rm K})$ and $\log_{10}(R_{\rm bb}/{\rm cm})$ using the affine-invariant MCMC sampler \texttt{emcee}, adopting uniform priors over $10^3<T_{\rm bb}<2\times10^5$\,K and $10^{13}<R_{\rm bb}<10^{17}$\,cm in logarithmic space.
The fitting results, with
$T_{\rm bb}=12{,}170^{+1{,}820}_{-1{,}340}$\,K,
$R_{\rm bb}=(9.08^{+1.67}_{-1.51})\times10^{14}$\,cm, and
$L_{\rm bb}=(1.29^{+0.28}_{-0.16})\times10^{43}$\,erg\,s$^{-1}$, are shown in the right panel of Figure~\ref{fig:opt_lc}. 

\subsection{Host Galaxy Analysis}\label{subsec:host}

\begin{figure}[htbp!]
    \centering
    \includegraphics[width=0.8\columnwidth]{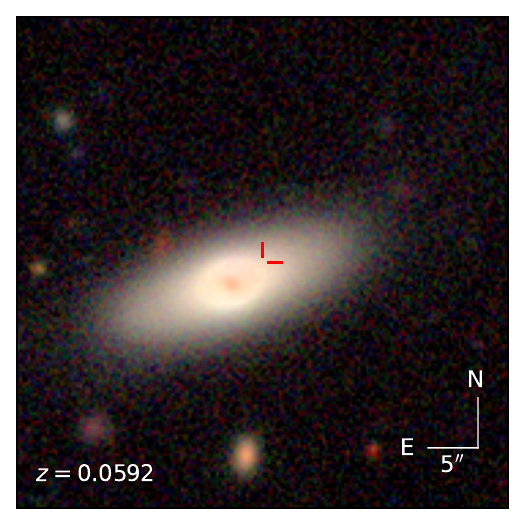}
    \caption{Legacy Survey RGB false-color $g$/$r$/$z$ image centered on the TDE candidate AT\,2024dky. The location of the transient is indicated with a red crosshair. The images were acquired between 2015 April and 2017 July. \label{fig:host_image} }
\end{figure}

AT\,2024dky is spatially associated with the a massive galaxy that was detected by multiple survey catalogs. Figure~\ref{fig:host_image} shows the optical archival image from the Legacy Survey (LS; \citealt{Dey2019}).
\citet{Dressler1980} initially designated it as MCG\,+03-38-036, classified it as an S0 galaxy and cataloged it as object No. 26 in the field of Abell 1983. \citet{Dressler1988_catalog} obtained spectroscopic redshift for candidate members of galaxy clusters. The redshift of MCG\,+03-38-036, $z=0.059154 \pm 0.000150$, lies well above the cluster median redshift of $z\approx0.045$. This indicates that it is most likely a background galaxy rather than a physical cluster member.
The galaxy nucleus was also observed by the sloan digital sky survey (SDSS; \citealt{Ahumada2020}) with a redshift of $z=0.059187\pm0.000013$ and a velocity dispersion of $\sigma_\ast = 273.2\pm6.3\,{\rm km\,s^{-1}}$. 
Hereafter we adopt the SDSS redshift for the host galaxy and the transient. 

The host-galaxy nucleus is located at 
${\rm R.A.}=14^{\rm h}51^{\rm m}17.808^{\rm s}$, 
${\rm decl.}=+16^{\circ}25^{\prime}59.690^{\prime\prime}$ in Gaia DR3 \citep{Gaia2023_dr3}.
The Gaia DR3 position of the host-galaxy nucleus differs
from the LS, SDSS, and Pan-STARRS (PS; \citealt{Chambers2016a}) catalog positions by
$0.015\arcsec$, $0.080\arcsec$, and $0.248\arcsec$,
respectively (see Figure~\ref{fig:offset}).
Using the Gaia positions of the transient and nucleus, we measure an angular offset of $3.724\arcsec$, which corresponds to a projected separation of $4.26$\,kpc.

Archival SDSS bulge--disk decompositions yield a bulge effective radius of $R_{\rm e,b}=1.39$\,kpc, an exponential disk scale length of $R_{\rm d}=4.65$\,kpc, and a $r$-band bulge-to-total light ratio of $B/T=0.57$ when the bulge S\'ersic index is fixed at $n=4$ \citep{Simard2011}. Allowing the bulge index to vary gives $n_{\rm b}=4.17$, $R_{\rm e,b}=1.56$\,kpc, $R_{\rm d}=4.93$\,kpc, and $B/T=0.60$. The high bulge S\'ersic index and substantial bulge contribution, together with the high stellar velocity dispersion, favor a classical bulge \citep{Kormendy2004}.
The corresponding \citet{Mendel2014} catalog also supplies component stellar masses, with ${\rm log}_{10}(M_{\rm bulge}/M_\odot) = 10.846^{+0.079}_{-0.150}$, ${\rm log}_{10}(M_{\rm disk}/M_\odot)=10.502^{+0.110}_{-0.160}$, and total galaxy mass of ${\rm log}_{10}(M_{\rm gal}/M_\odot)=10.953^{+0.113}_{-0.176}$. 

To obtain uniformly analyzed host-galaxy properties, we also performed our own spectral energy distribution (SED) fitting  to obtain stellar mass and star-formation rate (SFR). We model the host-galaxy photometry of AT\,2024dky with \texttt{prospector} \citep{Johnson2021}, following the procedures outlined in \citealt{Yao2025_24tvd}. From the joint posterior samples, we derive the surviving stellar mass, $M_{\rm gal}$, and the SFR averaged over the preceding 100\,Myr. For AT\,2024dky, we obtain $\log_{10}(M_{\rm gal}/M_\odot)=11.10^{+0.04}_{-0.07}$ and a model-dependent 95\% upper bound of $\mathrm{SFR}<0.061\,M_\odot\,\mathrm{yr}^{-1}$. These measurements are consistent with values reported in previous catalogs \citep{Mendel2014, Salim2016}.
We also repeat the same procedure for the hosts of the five known offset TDEs.

\subsection{Keck Late-time Imaging} \label{subsec:lris}

On 2026 June 18 (MJD 61210.4; $\delta t=858$\,d), deep $u^\prime$- and $R$-band images were acquired with the Low-Resolution Imaging Spectrometer (LRIS; \citealt{Oke1995}) on the Keck-I telescope, with $3\times5$\,min of exposure time in each filter (PI: R. Chornock). 
The observation was coordinated with the \textit{fritz.science} instance of \texttt{SkyPortal} \citep{vanderWalt2019, Coughlin2023}.
LRIS data were reduced using \texttt{lpipe} \citep{Perley2019lpipe}. 

\begin{figure*}[htbp!]
\centering
\includegraphics[width=0.8\linewidth]{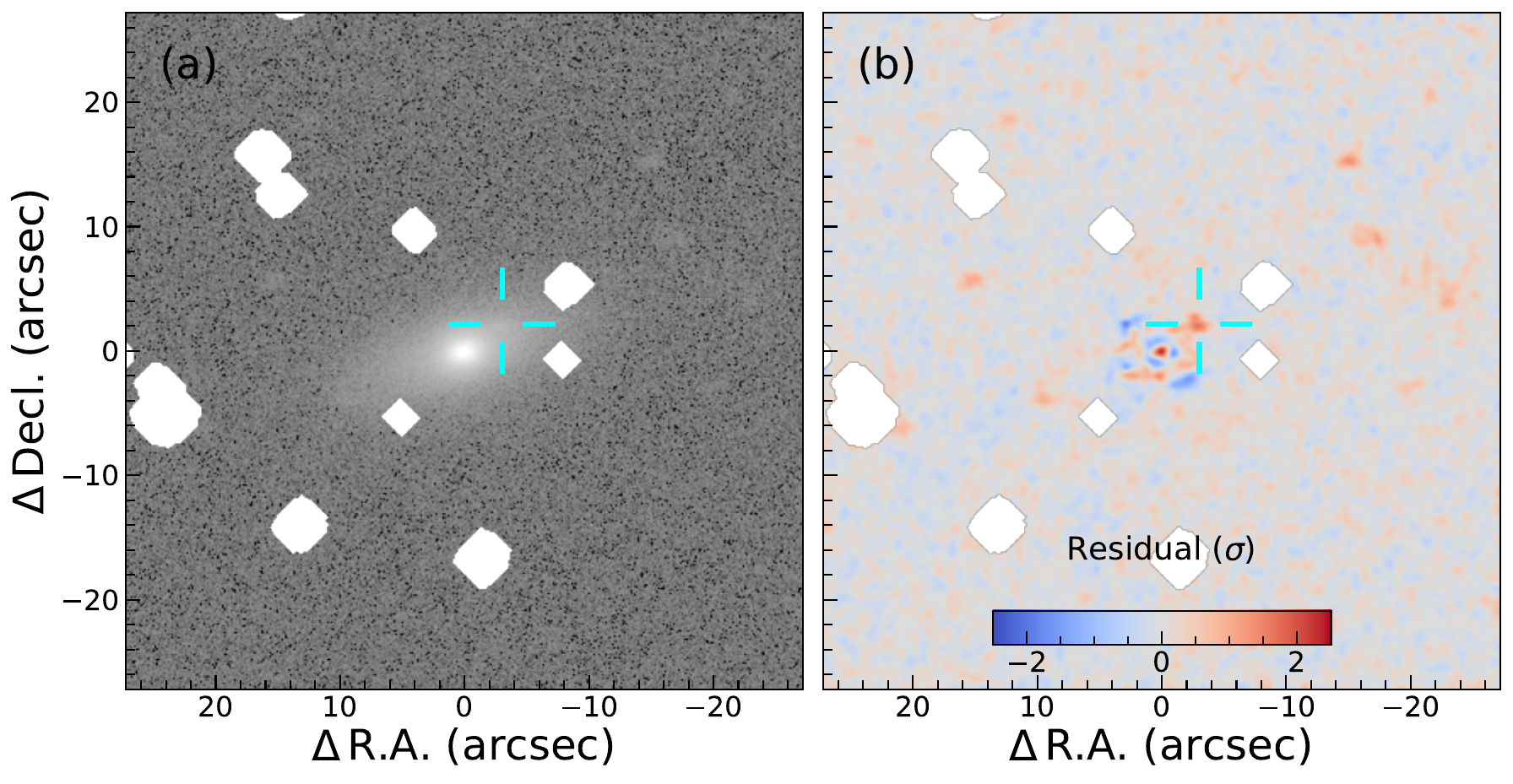}
\caption{Keck LRIS $u^\prime$ image (a) and residual after subtracting the \texttt{GALFIT} best-fit host galaxy model (b). The residual image is further smoothed with a Gaussian Kernel with the $\sigma$ of 2 pixel to display the transient marked by the cyan crosshairs. North is up and east is to the left.}
\label{fig:ubres}
\end{figure*}

To process the images for further analysis with \galfit\ \citep[][v3.0]{Peng2002, Peng2010}, we used the \texttt{photutils} package \citep{Bradley2020} to prepare the image cutouts, point spread function (PSF) models, and masks. 
For each frame, we first trimmed $\sim 150$--300 pixels from the edges that suffer from high noise and bad pixels. We used $3\sigma$ clipping to reject pixels containing source emission and estimate the background noise level. We then detected segments of emission above 1.5 times the noise level in the image convolved with a Gaussian kernel with a full width at half maximum (FWHM) of 3 pixels, requiring at least 5 pixels per segment. 
The detected segments were converted to source masks and expanded through 10 iterations of binary dilation to ensure that all source emission is covered by the masks.
The background is modeled with \texttt{Background2D} on a $15 \times 15$ pixel mesh using a median estimator and a $3 \times 3$ median filter, with $3\sigma$ clipping and all detected sources masked.
The resulting background map was subtracted from each frame. 
We constructed the PSF empirically from field stars selected as isolated, round sources (ellipticity $< 0.25$) with estimated FWHMs of $\sim 0.5$\arcsec--1.0\arcsec\ in $u^\prime$ and $\sim 1.0$\arcsec--1.65\arcsec\ in $R$, within a restricted segment-flux range. This selection yields $\sim 10$--20 stars. 
We combined these stars with \texttt{EPSFBuilder} \citep{Anderson2000} to obtain the effective PSF in each band.

To calibrate the magnitude zero point, we cross-matched the detected point sources with SDSS stars within 3\arcmin\ of the field center using a matching radius of 1\arcsec, yielding about 15 stars in each band. The SDSS $g$, $r$, and $i$ magnitudes are transformed to Cousins $R$ following \citet{Jester2005}, $R = g - 0.59(g-r) - 1.09(r-i) - 0.23$. 
We used the SDSS $u$ magnitudes to calibrate the LRIS $u^\prime$ photometry. The magnitude zero point uncertainties are
about 0.08~mag in both the $R$ and $u^\prime$ bands. 
We cross-matched Gaia 
sources within $1\arcmin$ of the target, with RUWE $< 1.5$, with sources 
detected at SNR $> 3$ within $1\arcsec$ to register the astrometry of our 
images. We find that the astrometric systematic uncertainty is 
$\lesssim 0.01\arcsec$ in both directions.

We modeled the surface brightness with \galfit, using the central $50\arcsec \times 50\arcsec$ image cutouts with surrounding sources masked.
We started with a single \sersic\ component for the host galaxy and gradually added more components to improve the fit. 

In the \uprime\ band, the host galaxy was well fitted with an exponential disk, a bulge with a \sersic\ index of $n\approx 1.8$, and a faint extended substructure $\sim 2\arcsec$ to the northeast. 
The best-fit center of the bulge is at 
${\rm R.A.}=14^{\rm h}51^{\rm m}17.80^{\rm s}$,
${\rm decl.}=+16^{\circ}25^{\prime}59.78^{\prime\prime}$. 
The latter component likely arises from small-scale non-uniformity in the local attenuation, which is common for a disk galaxy as highly inclined as ours.
An additional point source was required at
${\rm R.A.}=14^{\rm h}51^{\rm m}17.59^{\rm s}$,
${\rm decl.}=+16^{\circ}26^{\prime}01.92^{\prime\prime}$ (Figure~\ref{fig:ubres}), with $u^\prime=24.09 \pm 0.09$~mag, including statistical and calibration uncertainties. 
The statistical uncertainty was estimated by re-running \galfit\ fitting 50 times with the background level varying randomly according to the measured background fluctuation level. 
Comparing the best-fit results with and without the point source using the F-test, we find the additional point source is detected at a $11.8\sigma$ significance level.
After correcting for the Galactic extinction, we obtain $\uprime=23.99 \pm 0.09$~mag. 

In the $R$ band, the host galaxy was fitted with an exponential disk, a bulge with $n\approx 1.7$, and an intermediate size component with $n \approx 1$. 
The best-fit center of the bulge is at
${\rm R.A.}=14^{\rm h}51^{\rm m}17.80^{\rm s}$, 
${\rm decl.}=+16^{\circ}25^{\prime}59.82^{\prime\prime}$. 
The residual display moderate non-asymmetric pattern. 
We therefore quantified it with the Fourier mode on the exponential disk component. 
We found only the $m=2$ mode show a moderate amplitude of $\sim 0.15$. 
We did not find strong evidence of a recent merger based on the host galaxy morphology of the deep $R$~band image.  
We did not detect obvious emission at the location of the transient above the host galaxy residuals.
We therefore estimated an upper limit on the point-source flux at the position measured in the \uprime\ band. 
Starting from the best-fit two-component host model, we added a PSF component with its position and flux fixed, gradually increasing its flux, and refitting the other model parameters. 
The $\chi^2$ initially decreased as the PSF component absorbs some extended residual emission, but increased at higher PSF fluxes. 
We adopted a $3\sigma$ upper limit of 22.96\,mag, where $\chi^2$ increases by 9 relative to the best-fit model without a PSF component. 
After correcting for the Galactic extinction, we obtain $R>22.91$~mag.


Our $u^{\prime}$-band point source position is offset from the Gaia and ZTF median transient positions by 0.24\arcsec and 0.22\arcsec, respectively, primarily along the declination axis.
The variation in the background level introduces only $\lesssim 0.01\arcsec$ uncertainty in the point source declination. The total statistical astrometric uncertainty of the point source is therefore $<0.02\arcsec$. 
However, the fitted host-galaxy nuclei in the $u^\prime$ and $R$ bands are also offset from the Gaia nucleus by $0.15\arcsec$ and $0.18\arcsec$, respectively (Figure~\ref{fig:offset}).
These differences indicate systematic astrometric effects at the $\sim0.2\arcsec$ level, potentially arising from image registration and host-galaxy modeling.
Allowing for these effects and the characteristic $\sim0.1\arcsec$ uncertainty in the transient position (\S\ref{sec:discovery}),
the $u^\prime$-band point source is positionally consistent with AT\,2024dky.

\subsection{Radio: VLA} \label{subsec:vla}

On 2026 June 27 (MJD 61218.2; $\delta t = 866$\,d), we observed the field of AT\,2024dky under program 25B-109 (PI I. Sfaradi) for one hour with the National Science Foundation (NSF) Karl G. Jansky Very Large Array (VLA). The VLA was in a transitional, A$\rightarrow$D configuration and the observation was conducted with the X-band receiver (with a central frequency of $10$\,GHz). Interleaved observations of J1446+1721 were used to calibrate the complex gains, and 3C286 was used as a bandpass and absolute flux calibrator. We used the Common Astronomy Software Applications (CASA; \citealt{CASA}) packages and the VLA calibration pipeline (v6.5.4.9) to flag and calibrate the data, the CASA tasks \texttt{TCLEAN} and \texttt{IMSTAT} to image the field of AT\,2024dky and to calculate the image rms, respectively. Due to the hybrid configuration and small number of antennas with short baselines which cause an elongated beam, we introduce a cutoff in $uv$ space and image only baselines with separation larger than $40 \, k\lambda$.

\begin{figure}[htbp!]
    \centering
    \includegraphics[width=\columnwidth]{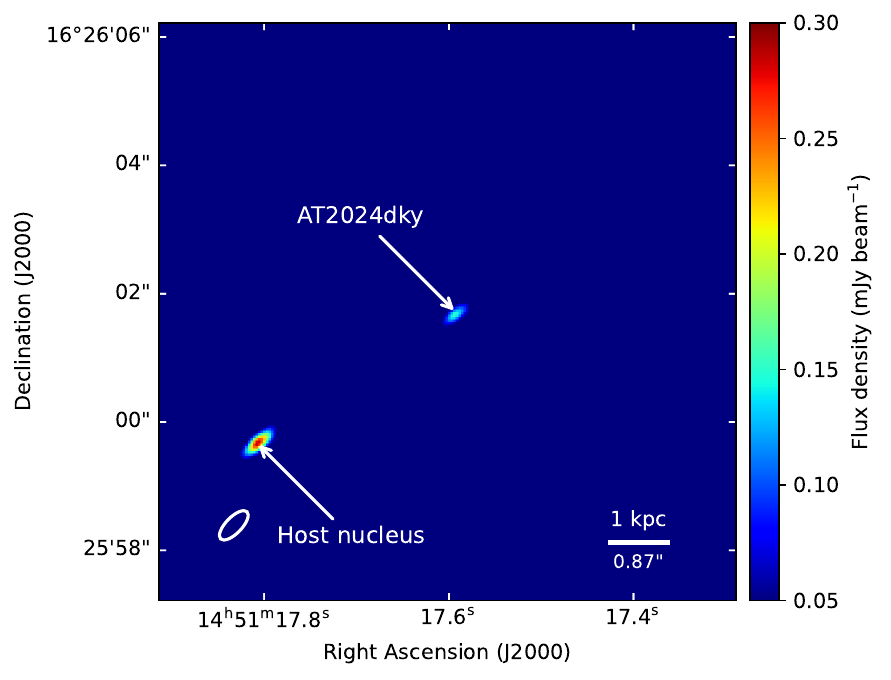}
    \caption{VLA X-band image in the field of AT\,2024dky. The white ellipse marks the VLA clean beam.\label{fig:radio_image}}
\end{figure}

We detect two point sources around the galactic center and fit both of them with the CASA task \texttt{IMFIT} (see Figure~\ref{fig:radio_image}). The first source is located at R.A.: 14$^{\rm h}$51$^{\rm m}$17$^{\rm s}$.8064 $\pm 0.0013 \, \rm s$, Dec.: +16$^{\rm \circ}$25$^{\rm '}$59$^{\rm "}$.693 $\pm 0.045^{\rm "}$, which is in agreement with the position of the center of the host. 
The second is at R.A.: 14$^{\rm h}$51$^{\rm m}$17$^{\rm s}$.5929 $\pm 0.0013 \, \rm s$, Dec.: +16$^{\rm \circ}$26$^{\rm '}$01$^{\rm "}$.680 $\pm 0.045^{\rm "}$, and is in agreement with the position of the optical transient (see Figure~\ref{fig:offset}). 
The positional uncertainties reported here correspond to $10\%$ of the full-width half-maximum of the synthesized beam. 
The flux density measurements of the source at the position of the transient is $F_{\rm \nu} = 0.15 \pm 0.02$\,mJy, corresponding to $\nu L_{\rm \nu} = \left( 1.26 \pm 0.17\right) \times 10^{38} \, \rm erg \, s^{-1}$. For the source at the center of the host we find $F_{\rm \nu }=0.28 \pm 0.03$\,mJy, which corresponds to $\nu L_{\rm \nu} = \left( 2.35 \pm 0.25 \right) \times 10^{38} \, \rm erg \, s^{-1}$. The uncertainties on the flux density reported here are a quadratic sum of the uncertainty provided by the best-fitting model, and additional $10\%$ calibration error.

\subsection{X-ray: EP} \label{subsec:ep}

On 2026 July 3 (MJD 61224.9; $\delta t = 872$\,d), we observed the field of AT\,2024dky with Follow-up X-ray Telescope (FXT; half energy width $\sim 30^{\prime\prime}$) on board the Einstein Probe (EP; \citealt{Yuan2022}) X-ray satellite for a total of 4495\,s. In the two FXT modules, nine counts were detected inside the 30$^{\prime\prime}$ radius aperture centered at the target position, whereas the expected number of background in the source aperture is 4.4 counts. This corresponds to a Poisson significance of only $1.8\sigma$, and we therefore do not claim a detection. Assuming a Galactic hydrogen column density \(N_{\rm H}=1.88\times10^{20}\ {\rm cm^{-2}}\) \citep{Dickey1990}, the 0.5--10\,keV flux upper limit are shown in Table \ref{tab:fxt} for different $\Gamma$. The upper limits were estimated independently for FXT-A and FXT-B, and the lower value was adopted.

\begin{table}[htbp]
\caption{\emph{EP}/FXT 90\% confidence upper limits on AT\,2024dky for different assumed power-law photon indices. \label{tab:fxt}}
\begin{tabular}{cccc}
\hline\hline
$\Gamma$ & $f_{\rm X,obs}$ & $f_{\rm X,unabs}$ & $L_{\rm X,unabs}$ \\
 & \multicolumn{2}{c}{(${\rm erg\,cm^{-2}\,s^{-1}}$)}
 & (${\rm erg\,s^{-1}}$) \\
\hline
2 & $1.02\times10^{-13}$ & $1.04\times10^{-13}$ & $8.71\times10^{41}$ \\
4 & $6.10\times10^{-14}$ & $6.53\times10^{-14}$ & $5.47\times10^{41}$ \\
6 & $6.70\times10^{-14}$ & $7.33\times10^{-14}$ & $6.14\times10^{41}$ \\
8 & $7.27\times10^{-14}$ & $8.04\times10^{-14}$ & $6.73\times10^{41}$ \\
10 & $7.39\times10^{-14}$ & $8.22\times10^{-14}$ & $6.88\times10^{41}$ \\
\hline
\end{tabular}
\tablecomments{Limits refer to the observed-frame 0.5--10\,keV band. The subscripts ``obs'' and ``unabs'' denote fluxes before and after correcting for Galactic absorption, respectively.
$L_{\rm X, unabs}$ does not strongly depend on $\Gamma$.}
\end{table}

\subsection{X-ray: Chandra} \label{subsec:cxo}

On 2026 September 15 (MJD 61298.7; $\delta t = 946$\,d), AT2024dky was observed with the Chandra X-ray Observatory using HRC-I under a DD request (PI: Y. Yao; ObsID 32461) for an exposure time of 9.43 ks. 
The data is contained in~\dataset[DOI: 10.25574/cdc.734]{https://doi.org/10.25574/cdc.734}.
The standard energy band for HRC-I is 0.1--10\,keV\footnote{\url{https://cxc.harvard.edu/csc/columns/ebands.html}}. This detector has poor energy resolution but good soft X-ray sensitivity\footnote{\url{https://cxc.harvard.edu/proposer/POG/html/chap7.html}}.

We reprocessed the data using CIAO v4.18 and CALDB v4.12.4 with \texttt{chandra\_repro}. 
We performed aperture photometry using \(1.5\arcsec\)-radius regions centered on the Gaia positions of the transient and the host-galaxy nucleus, which contained zero and eight events, respectively. The background was measured from a nearby source-free circular region with a radius of \(12\arcsec\). This region contained 101 events, corresponding to an expected background contribution of 1.58 events in each source aperture.

\begin{table}[htbp]
\caption{Chandra 90\% confidence upper limits on AT\,2024dky for different assumed power-law photon indices. \label{tab:chandra}}
\begin{tabular}{cccc}
\hline\hline
$\Gamma$ & $f_{\rm X,obs}$ & $f_{\rm X,unabs}$ & $L_{\rm X,unabs}$ \\
 & \multicolumn{2}{c}{(${\rm erg\,cm^{-2}\,s^{-1}}$)}
 & (${\rm erg\,s^{-1}}$) \\
\hline
2  & $5.70\times10^{-15}$ & $7.51\times10^{-15}$ & $6.30\times10^{40}$ \\
4  & $3.46\times10^{-15}$ & $2.14\times10^{-14}$ & $1.80\times10^{41}$ \\
6  & $3.18\times10^{-15}$ & $7.94\times10^{-14}$ & $6.66\times10^{41}$ \\
8  & $3.28\times10^{-15}$ & $2.57\times10^{-13}$ & $2.16\times10^{42}$ \\
10 & $3.72\times10^{-15}$ & $7.66\times10^{-13}$ & $6.43\times10^{42}$ \\
\hline
\end{tabular}
\tablecomments{All limits refer to the observed-frame 0.1--10\,keV
band. $L_{\rm X, unabs}$ strongly depends on $\Gamma$.}
\end{table}

AT\,2024dky was therefore not detected. Using \texttt{srcflux} and \texttt{aprates}, we obtained a 90\% confidence upper limit on its count rate of \(2.44\times10^{-4}\ {\rm counts\,s^{-1}}\). Assuming an absorbed power-law spectrum with different values of photon index \(\Gamma\) and Galactic column density, the corresponding 0.1--10 keV. flux upper limits are shown in Table~\ref{tab:chandra}. We note that the limit on unabsorbed $X$-ray luminosity, $L_{\rm X, unabs}$ strongly depends on the underlying spectral shape. 

\begin{table}[htbp]
\centering
\caption{Chandra X-ray flux and luminosity of the host-galaxy nucleus.}
\label{tab:cxo_nucleus}
\begin{tabular}{lccc}
\hline\hline
Band & Absorption & $f_{\rm X}$ & $L_{\rm X}$ \\
(keV) & & ($10^{-14}\,{\rm erg\,cm^{-2}\,s^{-1}}$)
        & ($10^{41}\,{\rm erg\,s^{-1}}$) \\
\hline
0.1--10 & Observed   & $1.58^{+0.80}_{-0.62}$ & $1.33^{+0.67}_{-0.52}$ \\
0.1--10 & Unabsorbed & $2.08^{+1.05}_{-0.82}$ & $1.75^{+0.88}_{-0.69}$ \\
2--10 & Observed & $0.73^{+0.37}_{-0.28}$ & $0.61^{+0.31}_{-0.24}$ \\
2--10   & Unabsorbed & $0.73^{+0.37}_{-0.29}$ & $0.61^{+0.31}_{-0.24}$ \\
\hline
\end{tabular}
\end{table}

For the host nucleus, the Poisson probability of detecting eight or more events from the background alone is \(2.4\times10^{-4}\), corresponding to a one-sided significance of \(3.5\sigma\). We therefore consider the host nucleus to be marginally detected in X-rays. Because the small number of events and the negligible spectral resolution of HRC-I preclude a meaningful spectral fit, we adopted an absorbed power-law model with $\Gamma = 2$ (typical for AGN; \citealt{Ricci2017}; see also \S\ref{subsec:center_smbh}) and obtained the flux and luminosity measurements as shown in Table~\ref{tab:cxo_nucleus}.

\section{The Nature of AT\,2024dky}

\begin{table*}[t]
\centering
\caption{Qualitative comparison of interpretations for AT\,2024dky.
}
\label{tab:classification}
\begin{tabular}{lcccccccc}
\toprule
Diagnostic & TDE  & SN IIn & SN Ibn/Icn & SN Ic-BL
           & SN IIb/Ibc & SN II & SN Ia \\
\hline
Persistently blue early optical color
 & $\checkmark$ & $\checkmark$ & $\checkmark$
 & $\times$ & $\times$ & $\times$ & $\times$ \\
Luminous late-time radio emission
 & $\checkmark$  & $\checkmark$ & $?$
 & $\checkmark$ & $\times$ & $\times$ & $\times$ \\
Massive, quiescent parent galaxy
 & $\checkmark$  & $?$ & $\times$
 & $\times$ & $\times$ & $\times$ & $\checkmark$ \\
Blue, luminous late-time optical emission
 & $\checkmark$ & $\times$ & $\times$
 & $\times$ & ? & $\times$ & ? \\
\bottomrule
\end{tabular}
\tablecomments{A check mark indicates consistency with the observed property; a cross indicates tension with observed examples; and a question mark indicates that the diagnostic is inconclusive. SLSNe and LFBOTs are omitted because they can be directly ruled out by the peak optical luminosity and duration.
}
\end{table*}

In this section, we assess the nature of AT\,2024dky by comparing its observed properties with those of TDEs and several classes of stellar transients. 

We note that its optical peak magnitude ($M_{\rm peak} = -18.5$\,mag) and duration ($t_{1/2} = 29.1$\,d) can directly rule out superluminous supernovae (SLSNe; $M_{g, \rm peak}<-19.8$\,mag; see \citealt{Gal-Yam2019} for a review) and luminous fast blue optical transients (LFBOTs; $M_{\rm peak}\lesssim-20$\,mag, $t_{1/2}<10$--15\,d; see \citealt{Ho2026} for a review), but are compatible with TDEs and all other types of SNe. 

Below, we compare AT\,2024dky with TDEs and SNe of types IIn, Ibn/Icn, Ic-BL, IIb/Ibc, II, and Ia. We examine its early-time optical emission (\S\ref{subsec:optical_peak}), late-time radio emission (\S\ref{subsec:radio_tde}), global host-galaxy properties (\S\ref{subsec:host_tde}), and late-time optical emission (\S\ref{subsec:optical_late}), which together strongly favor an off-nuclear TDE interpretation. Table~\ref{tab:classification} summarizes the diagnostic value of each property. 
Finally, we discuss the constraints from the X-ray non-detections (\S\ref{subsec:xray}).

\begin{figure*}[htbp!]
    \centering
    \includegraphics[width=0.95\textwidth]{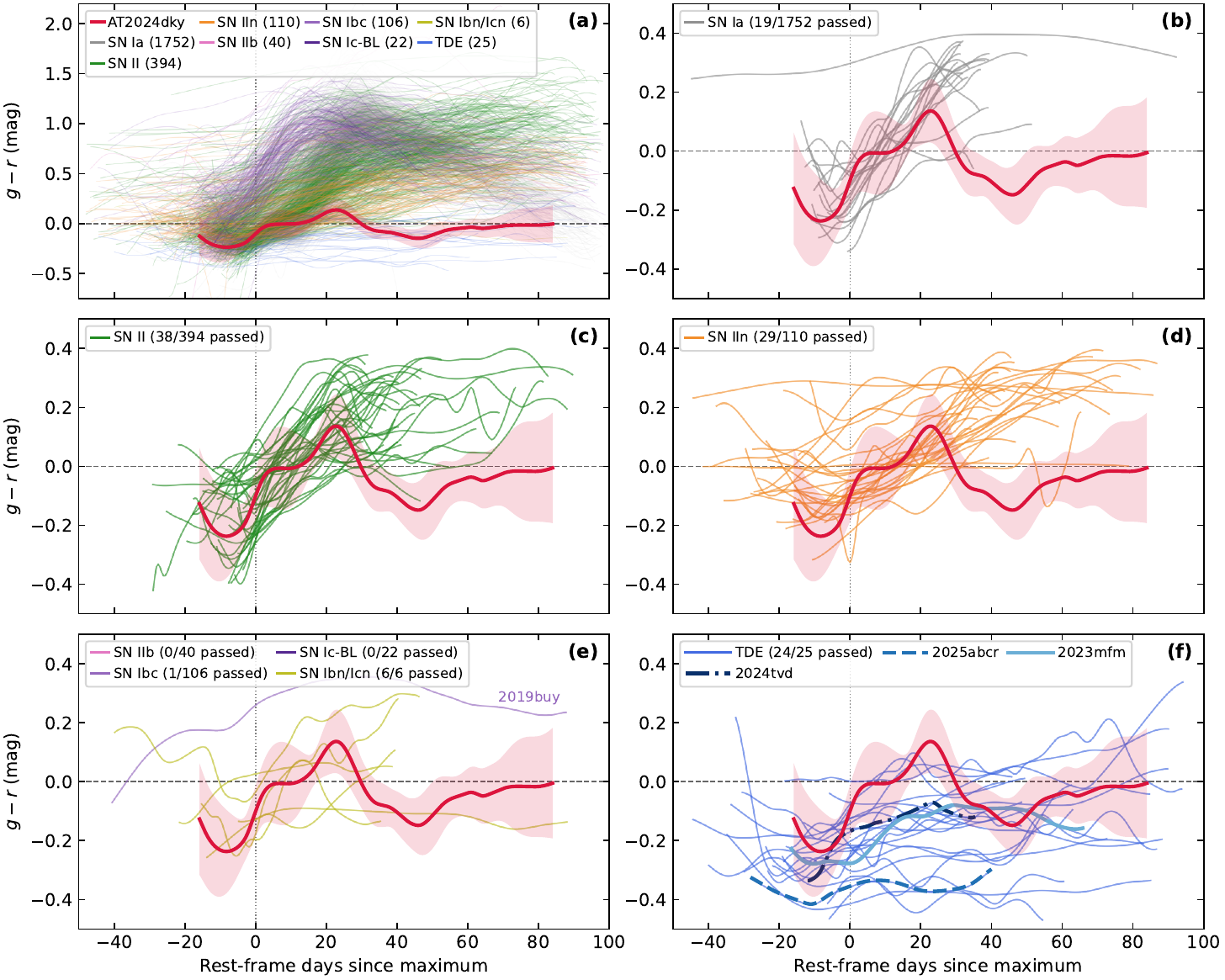}
    \caption{$g-r$ color evolution of AT\,2024dky and comparison transients with well-sampled ZTF light curves and $-20\leq M_{\rm peak}\leq-17$.
    The comparison light curves use ZTF alert-based photometry.
    Panel (a) shows the whole comparison sample. 
    Panels (b)--(f) show comparison objects separated by class, retaining only events whose GP colors satisfy $-0.5\leq g-r\leq0.4$ at every usable epoch between $-50$ and $+100$ rest-frame days. 
    Legends give the numbers passing this color selection relative to the samples in panel (a). 
    The three known off-nuclear TDEs are highlighted in panel (f). All three are in the comparison sample and passed our color criterion.
   \label{fig:opt_properties}}
\end{figure*}

\subsection{Optical Color Evolution} \label{subsec:optical_peak}

To compare the optical properties of AT\,2024dky with those of known transient classes, we construct a reference sample from the ZTF Bright Transient Survey (BTS; \citealt{Perley2020}) Sample Explorer\footnote{\url{https://sites.astro.caltech.edu/ztf/bts/explorer.php}}, selecting classified transients with $m_{\rm peak}<19.0$\,mag and peak times between 2019 January 1 and 2026 February 28. 
We fit multi-band GP models to the light curves following the method described in \S\ref{subsec:optical_peak}, using ZTF alert photometry queried through the \texttt{Kowalski} broker \citep{Duev2019}. 

We first use the full light curve to estimate the epoch of light curve maximum ($t_{\rm peak}$), and then only retain epochs within [$t_{\rm peak}-50$, $t_{\rm peak}+100$] to better characterize the light curve shape around peak.
We require observations in both $g$ and $r$ before $t_{\rm peak}$ and beyond $t_{\rm peak}+30(1+z)$, together with GP uncertainties below 0.5\,mag in both bands throughout the [$t_{\rm peak}$, $t_{\rm peak}+30(1+z)$] interval.
Restricting the comparison to events with measured redshifts and $-20\leq M_{\rm peak}\leq-17$\,mag (i.e., $\pm1.5$\,mag from the observed peak magnitude of AT\,2024dky) yields 2,455 objects. Their GP-derived $g-r$ evolution is shown in panel (a) of
Figure~\ref{fig:opt_properties}.
Most SNe redden substantially after maximum, whereas TDEs
generally retain blue, relatively constant colors.

We then select events whose $g-r$ colors satisfy $-0.5\leq g-r\leq0.4$\,mag across the fitted time range. This broad color cut identifies 139 events with blue colors broadly resembling those of AT\,2024dky. The fractions passing are 
$1.1^{+0.3}_{-0.2}\%$ for SNe~Ia (19/1,752),
$9.6^{+1.6}_{-1.4}\%$ for SNe~II (38/394),
$26.4^{+4.4}_{-4.0}\%$ for SNe~IIn (29/110),
$0.0^{+2.4}_{-0.0}\%$ for SNe~IIb (0/40),
$0.9^{+1.5}_{-0.6}\%$ for SNe~Ibc (1/106),
$0.0^{+4.4}_{-0.0}\%$ for SNe~Ic-BL (0/22),
$100.0^{+0.0}_{-14.3}\%$ for SNe~Ibn/Icn (6/6), and
$96.0^{+2.5}_{-6.0}\%$ for TDEs (24/25).
The quoted uncertainties represent 68.3\% Wilson binomial confidence intervals. 
We compare their color evolution with that of AT\,2024dky in panels (b)--(f) of Figure~\ref{fig:opt_properties}.

In stripped-envelope SNe of Types Ib, Ic, Ic-BL, and thermonuclear SNe Ia, the main optical peak is powered by radioactive decay, and the expanding ejecta subsequently cool.
In normal hydrogen-rich SNe II, hydrogen recombination can regulate the photospheric temperature, whereas in interacting SNe of Types IIn, Ibn, and Icn, continued heating through ejecta--circumstellar medium (CSM) interaction can sustain blue emission. These processes contribute to diverse color evolution, which can resemble that of TDEs over some intervals.

AT\,2024dky remains blue and exhibits no significant color evolution through 80 rest-frame days after maximum.
Panels (b)--(f) of Figure~\ref{fig:opt_properties} show that, even among events passing the broad color cut, the SNe Ia, SNe II, and the single SN Ibc generally become redder than AT\,2024dky beyond 30 rest-frame days after maximum. In contrast, a decent fraction of TDEs and SNe Ibn/Icn, together with a few SNe IIn, exhibit late-time color evolution resembling that of AT\,2024dky. Its peak absolute magnitude and half-maximum rise and decline times overlap those of several transient classes, providing less discrimination than its color evolution.

We conclude that the early-time optical properties of AT\,2024dky strongly favor a TDE or SN Ibn/Icn interpretation, remain compatible with an SN IIn origin, and strongly disfavor normal SNe Ia, SNe II, and SNe Ib/Ic/Ic-BL. 

\begin{figure}[htbp!]
    \centering
    \includegraphics[width=\columnwidth]{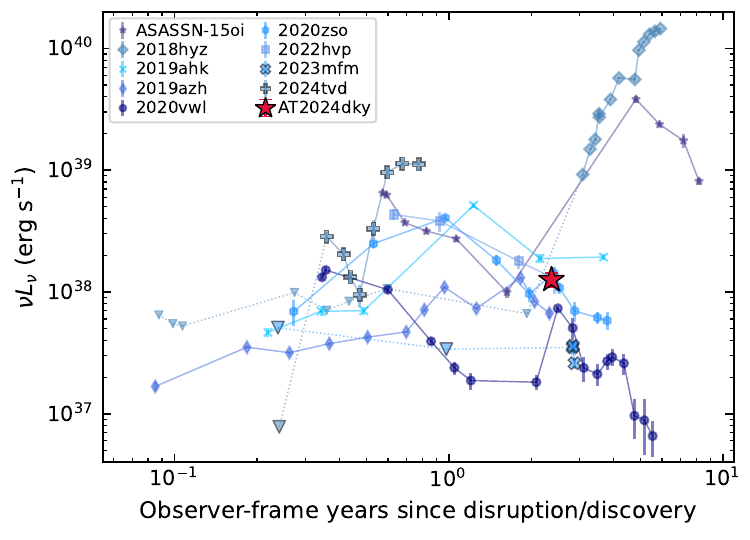}
    \includegraphics[width=\columnwidth]{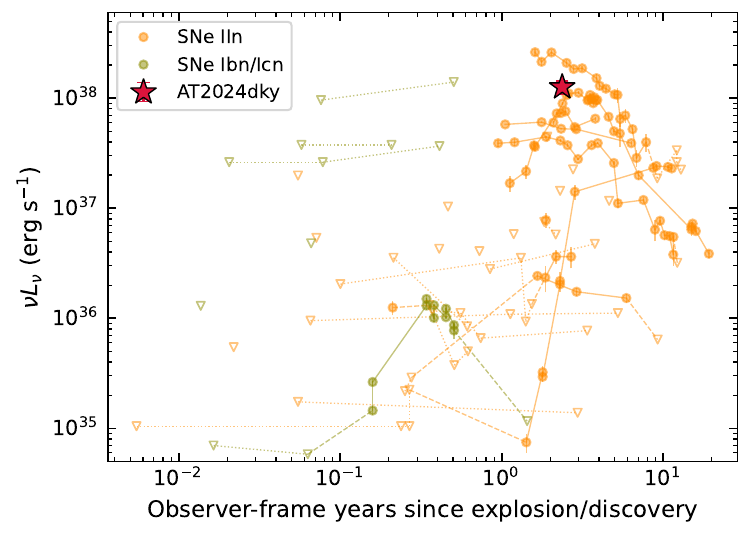}
    \caption{The $\sim 10$\,GHz radio luminosity of AT\,2024dky, compared with known TDEs (upper panel; \citealt{Goodwin2022, Goodwin2023_20vwl, Goodwin2025_20vwl, Hajela2025, Yao2025_radio, Sfaradi2024, Sfaradi2025, Sfaradi2026, Christy2024, Christy2026, Cendes2022, Cendes2026, Mummery2026_20vwl}, and interaction powered SNe (lower panel, see text for references).\label{fig:radio_lc} }
\end{figure}

\subsection{Late-time Radio Detection} \label{subsec:radio_tde}

At $\delta t=866$\,d, AT\,2024dky has a 10\,GHz radio luminosity of $\nu L_{\nu}=1.26\times10^{38}\,{\rm erg\,s^{-1}}$.
We compare this measurement with radio observations of SNe and TDEs. In both classes, the radio emission is generally attributed to synchrotron radiation from shocks driven by ejecta or outflows into the surrounding medium \citep[e.g.,][]{Alexander2020,Kundu2026}.

For core-collapse SNe (CCSNe), we use the compilation of \citet{Bietenholz2021}, which contains 1,475 flux-density measurements and upper limits for 294 SNe, drawn from the literature and from new and archival observations. The measurements are mostly at 4--10\,GHz and span early epochs through several years after explosion.
At its observed phase, AT\,2024dky lies at the extreme bright end of the observed distribution for SNe~II, IIb, and ordinary SNe~Ibc. Comparable or higher luminosities have been observed at late times in some SNe~IIn and Ic-BL \citep[][Figs.~3--7]{Bietenholz2021}.

Because SNe~Ibn and Icn are sparsely represented in the
\citet{Bietenholz2021} compilation, we supplement it with observations from \citet{Perley2022, Ho2023, Baer-Way2025, Stein2025, Wiston2026}.
The lower panel of Figure~\ref{fig:radio_lc} shows the resulting comparisons with interacting SNe. AT\,2024dky is about three orders of magnitude more luminous than the 10\,GHz upper limit for the SN~Ibn SN\,2023fyq at 525\,d after explosion \citep{Baer-Way2025}, which is the only Ibn/Icn event in our compilation with radio coverage beyond one year. This comparison provides limited leverage on the class as a whole.

For thermonuclear SNe, despite decades of targeted radio searches, the only confirmed SN~Ia with a late-time radio detection is the helium-rich SN~Ia-CSM SN\,2020eyj \citep{Kool2023, Griffith2025, Kundu2026}. 
It was detected at 5.1\,GHz roughly 600 and 740\,d after first optical detection with $\nu L_\nu \lesssim 10^{37}\,{\rm erg\,s^{-1}}$ \citep{Kool2023}, more than an order of magnitude below the 10\,GHz luminosity of AT\,2024dky.

The luminosity of AT\,2024dky is comparable to those of several
optically selected TDEs with long-term radio follow-up (upper panel of Figure~\ref{fig:radio_lc}). In a sample of 23 optically discovered TDEs, \citet{Cendes2024} found nine new radio detections in observations obtained $\sim500$--3200\,d after discovery, with late-time luminosities of order $10^{37}$--$10^{39}\,{\rm erg\,s^{-1}}$.

The late-time radio luminosity of AT\,2024dky is therefore consistent with a TDE, a radio-luminous SN~IIn, or an SN~Ic-BL. It disfavors ordinary SNe~Ia, II, IIb, and Ibc, while the sparse late-time coverage of SNe~Ibn/Icn leaves their radio luminosities at comparable phases poorly constrained.

\subsection{Host Galaxy Global Properties} \label{subsec:host_tde}

We compare the global stellar mass and SFR of AT\,2024dky's host with those of transient host populations to assess the host-environment diagnostics summarized in Table~\ref{tab:classification}. We focus on SNe~IIn/Ibn/Icn, which remain the principal alternatives after considering the early-time optical and late-time radio observations (\S\ref{subsec:optical_peak}--\ref{subsec:radio_tde}). The properties of AT\,2024dky's host and the five off-nuclear TDE comparison galaxies were derived uniformly using the analysis described in \S\ref{subsec:host}.

We construct a comparison sample of host galaxies of SNe IIn, Ibn, and Icn using published stellar-mass and SFR measurements. For SNe IIn and Ibn, we begin with the compilation of \citet{Schulze2021}, supplementing the Ibn sample with measurements from \citet{Sanders2013, Shivvers2016, Vallely2018, Kool2021, Warwick2025, Dong2026}. For SNe Icn, we adopt published host-galaxy measurements from \citet{Gal-Yam2022, Perley2022, Davis2023}; where a published mass–SFR pair is unavailable, we use the global host-galaxy SED-fitting results from the public BLAST database \citep{Jones2026}. We exclude measurements for which the SFR uncertainty exceeds 1.5 dex. The resulting comparison sample contains 105 SNe IIn, 14 SNe Ibn, and seven SNe Icn hosts. 

\begin{figure}[htbp!]
    \centering
    \includegraphics[width=\columnwidth]{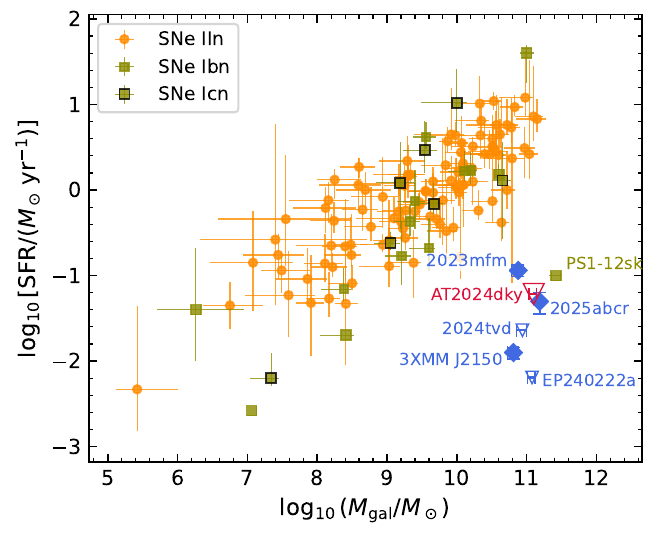}
    \caption{Host-galaxy SFR vs. stellar mass for AT\,2024dky, the parent galaxies of five comparison off-nuclear TDEs, and SNe~IIn/Ibn/Icn from the literature (see text). Downward triangles indicate 95\% SFR upper bounds. The global properties of AT\,2024dky's parent host galaxy bear remarkable similarities to those of known off-nuclear TDEs. \label{fig:host_mass_sfr}}
\end{figure}

As shown in Figure~\ref{fig:host_mass_sfr}, AT\,2024dky and the five comparison off-nuclear TDEs occupy a similar region of the mass--SFR plane, characterized by massive galaxies with low inferred SFRs or stringent upper bounds. 
In contrast, most SNe IIn, Ibn, and Icn are hosted by lower mass galaxies. 
All six SN~IIn hosts within 0.3\,dex of the stellar mass of AT\,2024dky's host have central SFR estimates of $\approx2.6$--$12\,M_\odot\,\mathrm{yr}^{-1}$, more than a factor of 40 above our upper bound of $0.061\,M_\odot\,\mathrm{yr}^{-1}$. 
The Ibn and Icn hosts generally follow the star-forming population, although an outlier is the massive, low-SFR host of PS1-12sk \citep{Sanders2013}. Thus, the massive, quiescent host of AT\,2024dky strongly supports an off-nuclear TDE interpretation and disfavors an interacting-SN origin.

For completeness, the massive, quiescent host also disfavors the other core-collapse SN classes considered in Table~\ref{tab:classification}, including SNe~Ic-BL, IIb/Ibc, and II, whose young progenitors are associated with recent star formation \citep{Schulze2021}. By contrast, such an environment is compatible with SNe~Ia \citep{Sullivan2006, Smith2012}. 

\subsection{Late-time Optical Properties} \label{subsec:optical_late}
We examine the late-time LRIS imaging results (\S\ref{subsec:lris}). 
At $\delta t = 858$\,d, the $u^\prime=23.99 \pm 0.09$\,mag detection and $R> 22.91$ upper limit gives $u^\prime - R <1.08$\,mag. 
The measurements correspond to 
$M_{u^\prime} = -13.07$\,mag or $\nu L_\nu = (6.74\pm0.56)\times 10^{40}\,{\rm erg\,s^{-1}}$,
and $M_R > -14.15$\,mag or $\nu L_\nu < 9.80\times 10^{40}\,{\rm erg\,s^{-1}}$, respectively.

The source is bluer and more luminous in the near-UV than globular clusters (GCs) and ultra compact dwarf (UCD) galaxies \citep{Peacock2010,Liu2015,Saifollahi2021}. The brightest M31 GC reaches $M_{u}\simeq-8.8$\,mag, while the brightest UCDs in the M87 and Fornax samples reach $M_{u^*}\simeq-10.9$\,mag and $M_{u}\simeq-11.8$\,mag, respectively. These populations have colors of $(u-r)>1$\,mag, disfavoring an old stellar system as the origin of the detected emission.

In hydrogen-rich interacting SNe, including SNe~IIn and SNe~II that develop strong interaction signatures at late times, continued ejecta--CSM interaction can power long-lasting optical emission, and the source is normally much brighter in a filter that covers H$\alpha$. 
For comparison, our GP fits to the long-term photometry of two SNe IIn presented by \citet{Stritzinger2012} yield $M_u=-13.53\pm0.40$~mag and $M_r=-15.75\pm0.38$~mag for SN\,2005ip ($u-r=2.22\pm0.55$\,mag), and $M_u=-14.69\pm0.24$~mag and $M_r=-16.67\pm0.22$~mag for SN\,2006jd ($u-r=1.98\pm0.33$\,mag), at the same phase. 
The LRIS $R$ filter extends farther redward than $r$ and encompasses H$\alpha$ at the redshift of AT\,2024dky.
Our $u^\prime-R$ color constraint therefore disfavors a SN IIn origin.

SNe~Ibn/Icn often exhibit rapidly declining optical light
curves, consistent with interaction involving radially
confined CSM or a steep outer density profile
\citep{Maeda2022}. 
Deep UV/optical imaging of two nearby SNe Ibn illustrates their faintness \citep{Sun2020}. SN\,2015G has limits of $M_{\rm F300X}>-6.36$\,mag and $M_{\rm F475X}>-5.31$\,mag at 1.8\,yr after peak. At 3--10\,yr, the source at the position of SN\,2006jc has $M_{\rm F300X}\simeq-5.00$\,mag and $M_{\rm F625W}\simeq-5.56$\,mag; its persistent emission is interpreted as arising from a surviving binary companion. Our $u^\prime$ detection therefore disfavors a SN Ibn/Icn origin. 

Strong late-time UV emission is also uncommon in SNe~Ia. 
An HST survey of 64 SNe~Ia detected late-time UV emission only from SN\,2015cp, which reached $M_{\rm F275W}\approx-13.1$\,mag at 664\,d after peak, attributed to delayed CSM interaction \citep{Graham2019_15cp}.
An optical spectrum obtained at $+694$\,d exhibits strong H$\alpha$ emission, whose measured flux corresponds to an absolute magnitude of $M_R\approx-11.8$\,mag.
Such emission would fall below our $R$-band detection limit; the non-detection therefore does not exclude a SN\,2015cp-like interaction episode.

For stripped-envelope SNe (SESNe), recent late-time HST observations of 91 objects yielded six secure UV detections: two SNe~Ic, two SNe~Ib, and two SNe~IIb, with $-14.4\leq M_{\rm F275W}\leq-10.8$\,mag at 765--1213\,d after optical discovery \citep{Fremling2026}.
These detections demonstrate that some SESNe can sustain substantial late-time UV emission, potentially powered by CSM interaction. Our LRIS measurements alone cannot exclude a similarly UV-luminous SESN.

For TDEs, the late-time emission of AT\,2024dky is consistent
with thermal emission from a persistent accretion disk \citep{Mummery2024_fundamental_scaling_relation,Guolo2025}.
Depending on the disk outer radius, the optical bands can sample the Rayleigh--Jeans regime ($f_\nu\propto\nu^2$; $u^\prime - R = -1.30$\,mag), the intermediate-frequency regime ($f_\nu\propto\nu^{1/3}$; $u^\prime - R = -0.22$\,mag), or the transition between them.

The late-time LRIS imaging therefore provides additional evidence favoring a TDE interpretation. We discuss the implications of the $u^\prime$-band detection for $M_{\rm BH}$ in \S\ref{subsec:Mbh}.

\subsection{Late-time X-ray Properties} \label{subsec:xray}
Late-time X-ray emission is expected for both SNe and TDEs. For SNe, the late-time X-ray comes from thermal bremsstrahlung or line emission produced by shock-CSM interaction. X-ray-luminous examples can reach $L_{\rm X}\sim10^{40}$--$10^{41}\,{\rm erg\,s^{-1}}$ and remain detectable for years, although their luminosities and evolution depend strongly on the CSM density structure and absorption \citep{AXIStdamm2024, Dwarkadas2025}. 
In TDEs, late-time X-rays can originate from the inner accretion disk. The luminosity in an observed X-ray band depends on the black hole mass, accretion rate, and disk viscous timescale, making a prediction at a given epoch difficult \citep{Mummery2025_fitted,Guolo2026_sample}.

Assuming an absorbed power-law spectrum with photon index $\Gamma=2$, our X-ray non-detections imply luminosity limits of
$L_{\rm X, unabs}<8.7\times10^{41}\,{\rm erg\,s^{-1}}$ in the 0.5--10\,keV band at $\delta t=872$\,d (\S\ref{subsec:ep}) and $L_{\rm X}<6.3\times10^{40}\,{\rm erg\,s^{-1}}$ in the 0.1--10\,keV band at $\delta t=946$\,d (\S\ref{subsec:cxo}). 
For the $\delta t=946$\,d Chandra observation, a softer intrinsic spectrum can allow a much higher intrinsic X-ray luminosity. 
The non-detections are compatible with both a TDE and a SN origin. 
In \S\ref{subsec:Mbh}, we examine how these limits constrain the allowed black hole masses under the accretion-disk interpretation.

\section{Discussion}
\subsection{Host Nucleus SMBH} \label{subsec:center_smbh}
A radio source with \(\nu L_{\nu,10\,\mathrm{GHz}}=(2.35\pm0.25)\times10^{38}\,\mathrm{erg\,s^{-1}}\) is detected at the host-galaxy nucleus (\S\ref{subsec:vla}). If powered entirely by star formation, adopting \(f_\nu\propto\nu^{-0.8}\) and the radio--SFR calibration of \citet{Davies2017} implies an SFR of $\sim 3.4\,M_\odot\,\mathrm{yr}^{-1}$. This exceeds our global SFR upper bound by a factor of $\sim 56$, strongly indicating that the nuclear radio emission is dominated by a radio AGN.

The Chandra detection of the host-galaxy nucleus implies a 2--10\,keV luminosity of $L_{\rm X, unabs}=6.1^{+3.1}_{-2.4}\times10^{40}\ {\rm erg\,s^{-1}}$ assuming a photon index of $\Gamma=2$ (see Table~\ref{tab:cxo_nucleus}). 
Adopting a radio spectral index of $\alpha=0.63$ ($f_\nu\propto\nu^{-\alpha}$; \citealt{Sabater2019}), the inferred 6\,cm (5\,GHz) radio luminosity is $\nu L_\nu=(1.82\pm0.19)\times10^{38}\ {\rm erg\,s^{-1}}$. Compared with AGN studies by \citet[][Fig.~2]{Panessa2007}, the host nucleus is a low-luminosity AGN (llAGN). 

Using the SDSS velocity dispersion (\S\ref{subsec:host}) and the \citet{Kormendy2013} $M_{\rm BH}$--$\sigma_\ast$ scaling relation, we obtained ${\rm log}_{10}(M_{\rm BH} / M_\odot) = 9.09\pm0.30$. If we instead use the \citet{Greene2020} $M_{\rm BH}$--$\sigma_\ast$ scaling relation calibrated with early-type galaxies, we obtained a similar result of ${\rm log}_{10}(M_{\rm BH} / M_\odot) = 9.02\pm0.44$. If we instead use the \citet{Kormendy2013} $M_{\rm BH}$--$M_{\rm bulge}$ scaling relation calibrated with classical bulges and elliptical galaxies, we obtained ${\rm log}_{10}(M_{\rm BH} / M_\odot) = 8.51^{+0.31}_{-0.34}\,M_\odot$.

Using the SDSS stellar velocity dispersion (\S\ref{subsec:host}), the $M_{\rm BH}$--$\sigma_\ast$ relation of
\citet{Kormendy2013} yields $\log_{10}(M_{\rm BH}/M_\odot)=9.09\pm0.30$.
The early-type galaxy calibration of \citet{Greene2020} gives a similar estimate of $\log_{10}(M_{\rm BH}/M_\odot)=9.02\pm0.44$.
Alternatively, applying the $M_{\rm BH}$--$M_{\rm bulge}$ relation of \citet{Kormendy2013} gives $\log_{10}(M_{\rm BH}/M_\odot)=8.51^{+0.31}_{-0.34}$.
Therefore, the galaxy center hosts a SMBH with a mass of ${\rm few}\times 10^8$--$10^9\,M_\odot$, powering a llAGN.

\subsection{Mass of the Candidate MBH Powering AT\,2024dky} \label{subsec:Mbh}

\begin{figure*}[htbp!]
    \centering
    \includegraphics[width=0.6\textwidth]{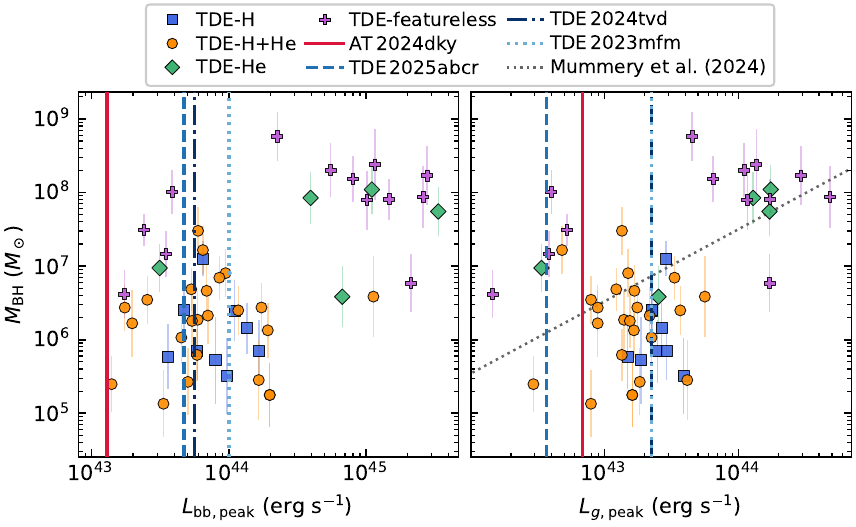}
    \caption{Black hole mass versus peak blackbody luminosity (left) and peak rest-frame $g$-band luminosity (right) for the ZTF nuclear TDE sample \citep{Yao2026}. $M_{\rm BH}$ is estimated from host galaxy scaling relations. 
Vertical lines mark the luminosities of AT\,2024dky and three previously known ZTF off-nuclear TDEs. \label{fig:L_Mbh}}
\end{figure*}

\subsubsection{From Spreading Disk Models}

At late times, once the accretion rate onto the black hole falls well below the Eddington rate, direct emission from a spreading accretion disk can account for the X-ray, UV, and optical emission of a TDE if reprocessing and shock emission have become subdominant (e.g., see \citealt{Cannizzo1990,vanVelzen2019_late_uv,Dai2021,Alush2025,Guolo2025}).
Theoretical models predict that the optical/UV plateau luminosity $L_{\rm plat}$ increases with black hole mass $M_{\rm BH}$, with the precise scaling depending on details of the models \citep{Mummery2024_fundamental_scaling_relation,Winter-Granic2026}.

Using a population of simulated relativistic TDE disks, \citet[][Equations~55--56]{Mummery2024_fundamental_scaling_relation} calibrated the $M_{\rm BH}$--$L_{\rm plat}$ relation at two rest-frame frequencies,
\begin{equation}
\log_{10}\left(\frac{M_{\rm BH}}{M_\odot}\right)
=1.50\log_{10}\left(\frac{L_{\rm plat}}{10^{43}\,{\rm erg\,s^{-1}}}\right)+b,
\end{equation}
where $L_{\rm plat}\equiv\nu L_{\nu,{\rm plat}}$, with $b=9.0$ at
$6\times10^{14}$\,Hz and $b=8.3$ at $10^{15}$\,Hz. 
If the late-time $u^\prime$-band detection of AT\,2024dky samples such a plateau, we can apply this relation at the corresponding rest-frame frequency, $\nu_{\rm rest}=9.2\times10^{14}$\,Hz. Retaining the common slope and interpolating the intercept linearly in $\log_{10}\nu$ gives $b\simeq8.41$. For the measured $\nu L_\nu=(6.74\pm0.56)\times10^{40}\,{\rm erg\,s^{-1}}$, we obtain $\log_{10}(M_{\rm BH}/M_\odot)=5.16\pm0.05\,({\rm stat})\pm0.75\,({\rm sys})$.

The X-ray non-detections provide a second constraint if emission from the inner disk emission is not absorbed or reprocessed. 
In Appendix~\ref{sec:fitted_model}, we jointly model the late-time Keck observations and X-ray upper limits with the relativistic spreading-disk model implemented in \texttt{FitTeD} \citep{Mummery2025_fitted}. The fit gives
$\log_{10}(M_{\rm BH}/M_\odot)=6.68^{+0.53}_{-0.55}$. 
The Chandra upper limit therefore disfavors the hotter low-$M_{\rm BH}$ solutions under the adopted assumptions. 

The fit also favors a small initial disk mass, with a posterior median of $M_{\rm disk}\simeq 1.7\times10^{-3}\,M_\odot$. This is difficult to reconcile with a full disruption in which a substantial fraction of the stellar debris forms the disk, calling into question whether the model adequately describes the late-time emission of AT\,2024dky (see more in \S\ref{subsubsec:Mbh_comment}).

\subsubsection{From Peak Optical Luminosity}

From the blackbody fit at $\delta t=0$\,d, we infer a rest-frame $g$-band peak luminosity of $\nu L_\nu=(6.81\pm0.21)\times10^{42}$\,erg\,s$^{-1}$.
Applying the empirical $L_{g,\rm peak}$--$M_{\rm BH}$ relation of \citet[][Equations 57, 74--77]{Mummery2024_fundamental_scaling_relation} gives $\log_{10}(M_{\rm BH}/M_\odot)=6.36\pm0.53$. The physical origin of this possible relation remains uncertain. 

\subsubsection{Comparison of Mass Estimates} \label{subsubsec:Mbh_comment}

The mass estimates above are subject to substantial uncertainties.

On one hand, the spreading-disk estimate is sensitive to how much inner disk emission escapes. Additional absorption or reprocessing could weaken the Chandra constraint on lower-mass models. 
For example, if a $\sim10^5\,M_\odot$ black hole were still accreting above or close to the Eddington rate 2--3\,yr after the flare, an optically thick outflow could redistribute their energy to longer wavelengths \citep{Dai2018,Thomsen2022_disk_wind}. 
The available observations leave this possibility open.

On the other hand, if lower-mass black holes produce intrinsically fainter optical flares, a flux-limited survey preferentially detects the luminous tail of their TDE population. This will bias the calibrated relation toward systematically overestimating the black hole masses of subluminous events. Quantifying the effect of this bias requires modeling both the intrinsic population and the survey selection function. 
Moreover, Figure~\ref{fig:L_Mbh} compares the peak luminosities and host-based black hole masses of 51 ZTF nuclear TDEs: the 49 comparison objects compiled by \citet[][\S4.1]{Yao2026}, together with the two TDEs studied in that work. Although the full sample exhibits a positive correlation over more than three orders of magnitude in $M_{\rm BH}$, this correlation is not evident within individual spectral classes. 

Nevertheless, we note that both the peak optical and late-time NUV luminosities of AT\,2024dky lie at the faint end of the observed TDE population.
Its peak blackbody luminosity is lower than that of all 51 nuclear TDEs in Figure~\ref{fig:L_Mbh}, and its peak rest-frame $g$-band luminosity is lower than those of 44 of the 51 objects (86\%).
Extrapolating the late-time $u^\prime$ luminosity to a rest-frame frequency of $10^{15}$\,Hz gives an NUV luminosity of $\nu L_\nu\simeq8\times10^{40}\,{\rm erg\,s^{-1}}$.\footnote{We obtain
$\nu L_\nu=(7.53\pm0.63)\times10^{40}\,{\rm erg\,s^{-1}}$ for
$f_\nu\propto\nu^{1/3}$ and
$(8.66\pm0.72)\times10^{40}\,{\rm erg\,s^{-1}}$ for
$f_\nu\propto\nu^2$.}
If the $u^\prime$ detection samples a plateau, this value is below the best-fit NUV plateau luminosities of 48 of the 49 TDEs in \citet[][Figure~8 and Table~E1]{Mummery2024_fundamental_scaling_relation}.

The low luminosities lead us to favor a black hole toward the low-mass end of the optically selected TDE population, with a characteristic mass of order $10^5$--$10^6\,M_\odot$.

\subsection{Origin of the Candidate MBH Powering AT\,2024dky} \label{subsec:bh_origin}

For AT\,2024dky, the existence of an AGN at the location of the parent galaxy's center, together with the low value of inferred $M_{\rm BH}$ powering the transient (\S\ref{subsec:Mbh}), rule out an origin in which the transient black hole is the galaxy's central SMBH displaced by gravitational-wave recoil.
Similar to other off-nuclear TDEs, the host galaxy image (Figure~\ref{fig:host_image}) and our deep $R$ band follow up image (\S\ref{subsec:lris}) do not show any sign of recent mergers (e.g., tidal features or tidal tails). Therefore, we were not able to definitively distinguish between other formation scenarios: minor merger that occurred long ago, dynamical ejection through a gravitational slingshot, and in-situ formation. 

The fact that the parent galaxies of AT\,2024dky and the other known off-nuclear TDEs have stellar masses clustered around $M_{\rm gal}\sim10^{11}\,M_\odot$ is interesting. 
As discussed by \citet{Yao2025_24tvd} for the three events then known, and subsequently examined by \citet{Guolo2026_wbh} using four events, this concentration is consistent with a population of wandering black holes deposited by disrupted satellite galaxies and retained at significant offset by long dynamical-friction timescales.
Their parent galaxies lie near the exponential cutoff of the local galaxy stellar mass function. This concentration could be qualitatively understood as a balance between the increasing abundance of wandering black holes per halo, which scales approximately linearly with dark matter halo mass \citep{Ricarte2021_wanderers_origin}, and the exponentially declining number density of more massive parent galaxies.

\subsection{Mass of the Stellar System Around the Candidate MBH}

We constrain the stellar mass of any compact stellar system associated with AT\,2024dky, such as a stripped galactic nucleus or a massive star cluster, using the late-time Keck/LRIS limit of $M_R>-14.15$\,mag.

Adopting $M_{R,\odot}=4.60$\,mag \citep{Willmer2018}, we obtain $L_R<3.2\times 10^7\,L_{R,\odot}$.
For an old stellar population, we adopt a stellar mass-to-light ratio of $\Upsilon_R=2.4\,M_\odot/L_{R,\odot}$ \citep[e.g.,][]{Penny2015}, yielding $M_{\star,\rm cluster}\lesssim7.6\times10^7\,M_\odot$.

\section{Summary}

We identified AT\,2024dky as a promising TDE candidate through a retrospective search of ZTF transients with persistently blue optical colors. It lies 3.72\arcsec\ (4.26\,kpc) from the center of an S0-type galaxy at $z=0.0592$. The parent galaxy hosts a central llAGN with a SMBH mass of ${\rm few}\times 10^8$--$10^9\,M_\odot$.

Our multi-wavelength observations establish the following
properties:
\begin{enumerate}
    \item A peak optical absolute magnitude of $M_{g,\rm peak}=-18.5$\,mag and a duration above half-maximum of $t_{1/2}=29.1$\,d.
    \item A persistently blue optical color through 80\,d after peak.
    \item A 10\,GHz radio luminosity of $\nu L_\nu=1.26\times10^{38}\,{\rm erg\,s^{-1}}$ at $\delta t=866$\,d.
    \item A massive parent galaxy ($M_{\rm gal}\sim10^{11}\,M_\odot$) with a low global SFR ($<0.061\,M_\odot\,\mathrm{yr}^{-1}$).
    \item A late-time detection at $M_{u^\prime}=-13.07\pm0.09$\,mag at $\delta t=858$\,d, with an $R$-band non-detection implying $u^\prime-R<1.08$\,mag.
    \item X-ray non-detections at $L_{\rm X}\lesssim 1.4\times10^{42}\,{\rm erg\,s^{-1}}$ ($\delta t=872$\,d) and $L_{\rm X}\lesssim 3\times10^{41}\,{\rm erg\,s^{-1}}$ ($\delta t=946$\,d).
\end{enumerate}
Together, these observations strongly favor an off-nuclear TDE interpretation for AT\,2024dky (Table~\ref{tab:classification}).

While the black hole mass is highly uncertain, we favor a low-mass MBH with $M_{\rm BH}\sim 10^5$--$10^6\,M_\odot$. 
Further multi-wavelength follow-up observations will constrain properties of the black hole and the surrounding stellar system.


\appendix

\section{Spreading Thin Disk Modeling of Late-time Emission} \label{sec:fitted_model}

\begin{figure*}[htbp!]
    \centering
    \includegraphics[width=\textwidth]{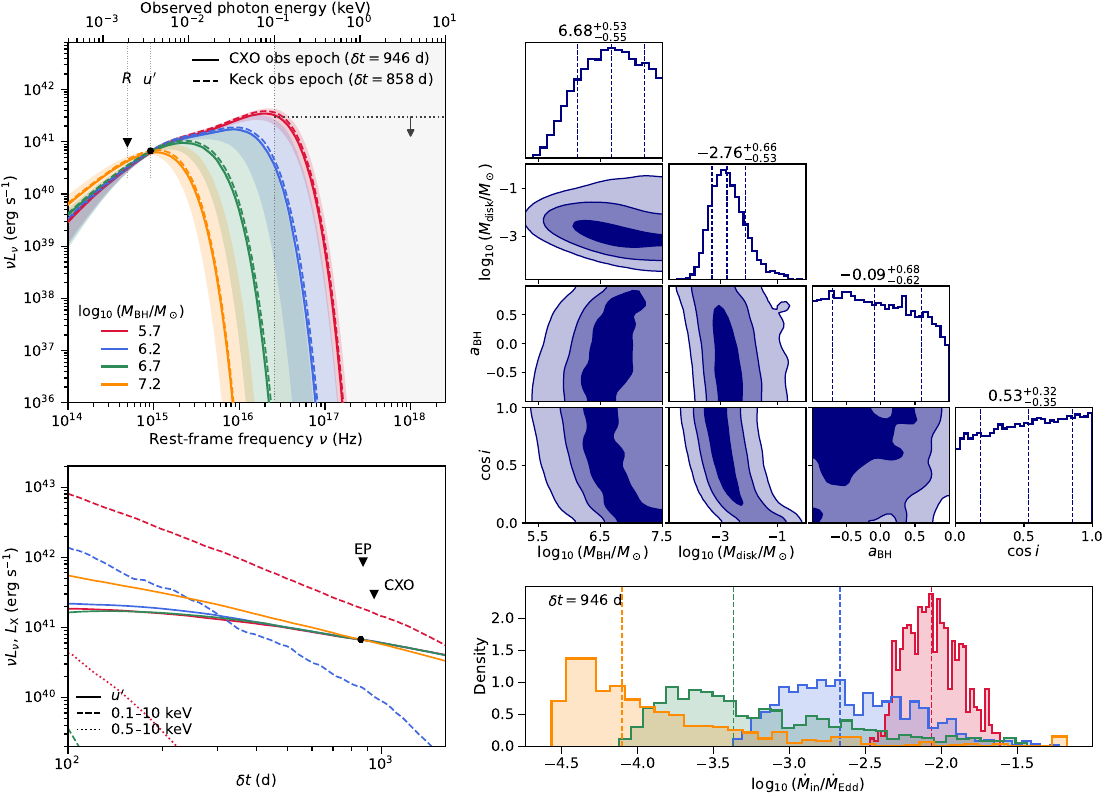}
    \caption{Spreading-disk modeling results with \texttt{FitTeD} for the late-time emission of AT\,2024dky.
    Colors identify posterior samples within \(\pm0.05\) dex of \(\log_{10}(M_{\rm BH}/M_\odot)=5.7,\ 6.2,\ 6.7,\) and \(7.2\). 
    \textit{Upper left}: Intrinsic SEDs at the Keck-I/LRIS  and Chandra observing epochs. Black symbols show the LRIS \(u'\) detection and \(R\)-band upper limit. 
    The horizontal dotted line marks the adopted 0.1--10\,keV Chandra limit, \(L_{\rm X}<3\times10^{41}\) erg s\(^{-1}\). 
    \textit{Lower left}: Median disk light curves in different bands, compared with the $u^\prime$-band detection and X-ray upper limits. 
    \textit{Upper right}: Posteriors of $M_{\rm BH}$, $M_{\rm disk}$, $a_{\rm BH}$, and ${\rm cos}i$, marginalized over $r_0$ and $t_{\rm visc}$. 
    \textit{Lower right}: Distributions of the inner accretion rate at the Chandra observing epoch, normalized by \(\dot M_{\rm Edd}=L_{\rm Edd}/(\eta c^2)\); dashed vertical lines mark medians. 
    \label{fig:fitted}}
\end{figure*}

Under the assumption that the late-time X-ray/UV/optical emission comes from direct emission of a spreading thin accretion disk, we use \texttt{FitTeD}\footnote{\url{https://fittingtransientswithdiscs.github.io/FitTeD/}} \citep{Mummery2025_fitted} to explore which spreading-disk models can reproduce the late-time $u^\prime$ detection while satisfying the X-ray upper limits. 

We use the \texttt{GR\_disc} model, which evolves an initially narrow ring of debris under relativistic thin-disk dynamics and computes a color-corrected thermal spectrum from the evolving radial temperature profile \citep{Mummery2025_fitted}. The disk model has seven parameters: the black hole mass $M_{\rm BH}$ and dimensionless spin $a_{\rm BH}$, the initial disk mass $M_{\rm disk}$ and radius $r_0$, the rest-frame viscous timescale $t_{\rm visc}$, the disk formation epoch $t_0$, and the viewing inclination $i$.

We fix $t_0$ at the optical maximum because it is well measured and occurs far earlier than the epochs used in the fit. 
We sample the six remaining parameters using MCMC, with independent uniform priors on $\log_{10}(M_{\rm BH}/M_\odot)$ over $[4,7.5]$,
$\log_{10}(M_{\rm disk}/M_\odot)$ over $[-5,0]$,
$\log_{10}(r_0/r_{\rm g})$ over $[1,4]$, and
$\log_{10}(t_{\rm visc}/{\rm d})$ over $[0,3.5]$, where
$r_{\rm g}=GM_{\rm BH}/c^2$. 
We adopt a uniform prior on $a_{\rm BH}\in[-0.998,0.998]$ and a uniform prior on $\cos i\in[0,1]$.

We restrict the fit to the late-time data, using a Gaussian likelihood for $M_{u^\prime}=-13.07\pm0.09$\,mag at $\delta t=858$\,d and treating the contemporaneous $R$-band limit of $M_R>-14.15$\,mag and the X-ray upper limits as hard bounds.
For EP, we adopt the 0.5--10\,keV luminosity limit of $8.7\times10^{41}\,{\rm erg\,s^{-1}}$ at $\delta t=872$\,d (\S\ref{subsec:ep}).\footnote{The EP luminosity limit also depends on the assumed spectrum and can be less stringent for a soft thermal disk spectrum. For the models explored here, the EP limit adds little constraint beyond the Chandra limit (Figure~\ref{fig:fitted}).}
For Chandra, we adopt a fiducial intrinsic luminosity limit of $L_{\rm X}<3.0\times10^{41}\,{\rm erg\,s^{-1}}$ in the observer-frame 0.1--10\,keV band at $\delta t=946$\,d. The conversion from the Chandra count-rate limit to an intrinsic luminosity depends on the assumed spectrum (\S\ref{subsec:cxo}). We will later assess this fiducial choice by folding posterior spectra through the HRC-I response.

Figure~\ref{fig:fitted} shows that models spanning a broad range of black hole masses reproduce the late-time $u^\prime$ luminosity. Their predicted X-ray luminosities differ by orders of magnitude, with the lowest-mass models shown approaching the adopted Chandra limit. At fixed $u^\prime$ luminosity, higher black hole masses generally require smaller initial disk masses and yield lower Eddington-scaled inner accretion rates at the Chandra observing epoch. Marginalizing over $r_0$ and $t_{\rm visc}$, both of which are weakly constrained, gives $\log_{10}(M_{\rm BH}/M_\odot)=6.68^{+0.53}_{-0.55}$ and $\log_{10}(M_{\rm disk}/M_\odot)=-2.76^{+0.66}_{-0.53}$. 
The disk fit favors a higher mass than the plateau relation under the adopted assumptions. 

To assess the adopted Chandra luminosity bound, we draw 10,000 spectra at the Chandra epoch from the full posterior, apply Galactic absorption, and fold each spectrum through the HRC-I response. The predicted count rate has a 95th percentile of $1.09\times10^{-4}\,{\rm counts\,s^{-1}}$, and only 0.29\% of the draws exceed the measured 90\% upper bound of $2.44\times10^{-4}\,{\rm counts\,s^{-1}}$. Thus, the adopted intrinsic 0.1--10\,keV luminosity bound of $3.0\times10^{41}\,{\rm erg\,s^{-1}}$ is a reasonable approximation for identifying models consistent with the Chandra non-detection.

The joint fit favors an initial disk mass of only $M_{\rm disk}\simeq1.7\times10^{-3}\,M_\odot$.
Only about 3\% of posterior draws reach $M_{\rm disk}\geq0.04\,M_\odot$, the minimum initial disk mass in the population models of \citet{Mummery2024_fundamental_scaling_relation}, who assume that half of the disrupted star's mass forms the disk. 
These higher disk masses occur predominantly at nearly edge-on inclinations.
Thus, within the adopted spreading-disk model, reproducing the late-time data is difficult if a substantial fraction of the debris from a full disruption settles into the initial disk.

\begin{acknowledgments}

EMR acknowledges support from European Research Council (ERC) grant number: 101002511/project acronym: VEGA P and EMR acknowledges that this publication is part of the project ``Flares from disrupted stars unveil the origin of the giant black holes'' with file number  VI.C.232.099 of the research programme VICI.

Based on observations obtained with the Samuel Oschin Telescope 48-inch and the 60-inch Telescope at the Palomar Observatory as part of the Zwicky Transient Facility project. ZTF is supported by the National Science Foundation under Grants No. AST-1440341, AST-2034437, and currently Award \#2407588. ZTF receives additional funding from the ZTF partnership. Current members include Caltech, USA; Caltech/IPAC, USA; University of Maryland, USA; University of California, Berkeley, USA; Cornell University, USA; Drexel University, USA; University of North Carolina at Chapel Hill, USA; Institute of Science and Technology, Austria; National Central University, Taiwan, German Center for Astrophysics, Germany, and OKC, University of Stockholm, Sweden. Operations are conducted by Caltech's Optical Observatory (COO), Caltech/IPAC, and the University of Washington at Seattle, USA. 

The ZTF forced-photometry service was funded under the Heising-Simons Foundation grant \#12540303 (PI: Graham).

The Gordon and Betty Moore Foundation, through both the Data-Driven Investigator Program and a dedicated grant, provided critical funding for SkyPortal.

Some of the data presented herein were obtained at Keck Observatory, which is a private 501(c)3 nonprofit organization operated as a scientific partnership among the California Institute of Technology, the University of California, and the National Aeronautics and Space Administration. The Observatory was made possible by the generous financial support of the W. M. Keck Foundation. 
The authors wish to recognize and acknowledge the very significant cultural role and reverence that the summit of Maunakea has always had within the Native Hawaiian community. We are most fortunate to have the opportunity to conduct observations from this mountain.

This research has made use of data obtained from the Chandra Data Archive provided by the Chandra X-ray Center (CXC).

The National Radio Astronomy Observatory (NRAO) is a facility of the National Science Foundation operated under cooperative agreement by Associated Universities, Inc. We thank the NRAO for carrying out the Karl G. Jansky Very Large Array (VLA).

This work has made use of data obtained with the Einstein Probe, a space mission supported by the Strategic Priority Program on Space Science of the Chinese Academy of Sciences, in collaboration with ESA, MPE and CNES.

RC, IS, and RM acknowledge the hospitality of the Kavli Institute for Theoretical Physics at UC Santa Barbara, where portions of this work were completed. This research was supported in part by the National Science Foundation under PHY-2309135.

\end{acknowledgments}

\bibliography{main}{}
\bibliographystyle{aasjournalv7.1}

\end{CJK*}
\end{document}